\documentclass[iop]{emulateapj}  

\newcommand\aov{\ifmmode{\alpha_{\rm ov}}\else $\alpha_{\rm ov}$\fi}
\newcommand\fov{\ifmmode{f_{\rm ov}}\else $f_{\rm ov}$\fi}
\newcommand\amix{\ifmmode{\alpha_{\rm MLT}}\else $\alpha_{\rm MLT}$\fi}
\newcommand\cb{CB2018}
\newcommand\na{New Astronomy}

\shortauthors{Claret \& Torres}
\shorttitle{Core overshooting}

\begin{document} 

\submitted{Accepted for publication in The Astrophysical Journal}

\author{
Antonio Claret\altaffilmark{1} and
Guillermo Torres\altaffilmark{2}
}

\altaffiltext{1}{Instituto de Astrof\'{\i}sica de Andaluc\'{\i}a,
  CSIC, Apartado 3004, 18080 Granada, Spain and \and Dept. F\'{\i}sica
  Te\'{o}rica y del Cosmos, Universidad de Granada, Campus de
  Fuentenueva s/n, 10871, Granada, Spain}

\altaffiltext{2}{Center for Astrophysics \textbar\ Harvard \&
  Smithsonian, 60 Garden St., Cambridge, MA 02138, USA}

\begin{abstract}

Overshooting from the convective cores of stars more massive than
about 1.2~$M_{\sun}$ has a profound impact on their subsequent
evolution. And yet, the formulation of the overshooting mechanism in
current stellar evolution models has a free parameter (\fov\ in the
diffusive approximation) that remains poorly constrained by
observations, affecting the determination of astrophysically important
quantities such as stellar ages. In an earlier series of papers we
assembled a sample of 37 well-measured detached eclipsing binaries to
calibrate the dependence of \fov\ on stellar mass, showing that it
increases sharply up to a mass of roughly 2~$M_{\sun}$, and remains
constant thereafter out to at least 4.4~$M_{\sun}$. Recent claims have
challenged the utility of eclipsing binaries for this purpose, on the
basis that the uncertainties in \fov\ from the model fits are
typically too large to be useful, casting doubt on a dependence of
overshooting on mass. Here we reexamine those claims and show them to
be too pessimistic, mainly because they did not account for all
available constraints --- both observational and theoretical --- in
assessing the true uncertainties. We also take the opportunity to add
semi-empirical \fov\ determinations for 13 additional binaries to our
previous sample, and to update the values for 9 others. All are
consistent with, and strengthen our previous conclusions, supporting a
dependence of \fov\ on mass that is now based on estimates for a total
of 50 binary systems (100 stars).
\end{abstract}


\title{The Dependence of Convective Core Overshooting on Stellar Mass: \\
  Reality Check, and Additional Evidence}

\section{Introduction}
\label{sec:introduction}

In recent years there has been a resurgence of interest in the
phenomenon of convective core overshooting in stars. Empirical
constraints of different kinds have been brought to bear on the
problem of calibrating the free parameters of the two most often used
prescriptions for overshooting, which are \aov\ for the classical
step-function implementation, and \fov\ for the diffusive
approximation \citep{Freytag:1996, Herwig:1997}. In the classical
formulation the extension of the core beyond the boundary set by the
Schwarzschild criterion is $d_{\rm ov} = \aov H_p$, whereas in the
alternate prescription the mixing is modeled as a diffusive process
with a diffusion coefficient at a radial distance $r$ from the
boundary given by $D(r) = D_0 \exp{(-2r/\fov H_p)}$, in which $D_0$ is
the coefficient inside the boundary and $H_p$ the pressure scale
height. In a series of papers over the last three years
\citep[][hereafter Paper~I, Paper~II, and Paper~III]{Claret:2016,
  Claret:2017, Claret:2018} we employed a total of 37 double-lined
eclipsing binaries (DLEBs) with well-measured masses, radii, effective
temperatures, and in some cases metallicities, to examine the
dependence of the strength of overshooting on stellar mass. The
measurements were compared against current stellar evolution models to
infer semi-empirical values for the overshooting parameter for each
star.

It was found that \aov\ and \fov\ increase sharply starting at
$\sim$1.2~$M_{\sun}$ to a maximum at $\sim$2~$M_{\sun}$, after which
they show little or no change up to the 4.4~$M_{\sun}$ limit of our
sample. The peak value for \fov\ is about 0.016. The mass behavior was
found to be independent of the element mixtures we tested
\citep{Grevesse:1998, Asplund:2009}, and was qualitatively the same
for \aov\ and \fov, except for a scaling factor between the two of
$\aov/\fov \approx 11.4$ (Paper~II).

In recent work \cite{CB:2018} (hereafter \cb) have revisited the issue
with the goal of assessing the degree to which observations for DLEBs
constrain the overshooting parameter. They focused specifically on
\fov, and selected a subsample of eight of our binaries for their
experiments, which they considered to be representative of the range
of stellar masses and evolutionary states of the parent population.
Their modeling led them to claim that in most cases the range of
\fov\ values permitted by the observations is very large (often the
full interval they explored, typically $\fov = 0.000$--0.040), such
that it is very difficult to discern any dependence at all on stellar
mass. They concluded therefore that the method of using DLEBs for this
type of calibration is of limited utility.  Overall they found that a
constant value of the overshooting parameter around 0.013 or 0.014
seems adequate to fit all eight of the systems they studied, given
that they have such large uncertainties. Values of \fov\ of this order
would be consistent with the degree of overshooting we found in our
previous work for masses larger than about 2~$M_{\sun}$ (Paper~II,
Paper~III), but not for lower mass stars.

Intrigued by the \cb\ claims that seem to conflict with our own
experience, as a first goal we set out here to examine their results
more closely, and to repeat their experiments following their
grid-based approach to the extent possible. We depart from their basic
procedure only in that we use the same stellar evolution models and
physical ingredients that we employed in our previous series of
papers, to ensure consistency with our earlier results. To our
surprise we find significant disagreements with \cb\ for most of the
systems, which in some cases may be explained by the differences in
the models.  Furthermore, we believe their assessments are overly
pessimistic in part because they did not consider other available
constraints, some empirical and some rooted in theory, that
significantly reduce the formal uncertainties in \fov\ for some
systems. In particular, we review physical arguments that set a
maximum size for the convective cores of stars in the low- and
intermediate-mass regime ($\lesssim 2~M_{\sun}$), effectively making
values of \fov\ as large as some of those proposed by
\cb\ unrealistic. We believe these arguments, along with our own grid
experiments for the eight binaries studied by \cb, show that in many
cases DLEB observations do indeed have enough discriminating power to
discern a variation of \fov\ with mass.

As a second goal of this paper we take the opportunity to report
semi-empirical determinations of \fov\ for an additional set of
eclipsing binaries we have identified that are suitable for this type
of analysis. As we will show, those results are consistent with and
strengthen the general trend found earlier.

We have organized our paper as follows. In
Section~\ref{sec:methodology} we summarize our procedures for
exploring the range of acceptable values of \fov\ for the systems
studied by \cb. Descriptions of these grid experiments are presented
and discussed individually in Section~\ref{sec:grids} along with a
comparison against the results \cb, highlighting the disagreements.
Our conclusions regarding the claims by \cb\ are given in
Section~\ref{sec:fovstrength}.  In Section~\ref{sec:newbinaries} we
expand our previous sample of 37 binaries and determine semi-empirical
values of \fov\ for another 13 DLEBs. We also revise the values for 9
of them, based on improvements to their stellar parameters that have
appeared recently in the literature. This section also reports our
updated \fov\ vs.\ mass relation in a new rendering that more fairly
displays the uncertainties in our determinations. We conclude with a
discussion in Section~\ref{sec:discussion}.  Appendix~\ref{sec:theory}
describes in some detail important physical arguments suggesting that
the degree of overshooting cannot be as large as claimed by \cb\ for
stars of low and intermediate mass.

\section{Grid search methodology}
\label{sec:methodology}

\cb\ based their study on eight DLEBs listed in their Table~1. Three
have component masses smaller than 2~$M_{\sun}$ and are therefore of
particular interest (AY~Cam, HD~187669, and BK~Peg), and the other
five are more massive. In most cases the primary and secondary stars
are fairly similar in terms of their mass. For each binary they
explored a range of values of the overshooting parameter \fov, the
standard mixing-length parameter \amix, and the system metallicity
[Fe/H], all of which they report in the same table.  They assumed both
\amix\ and \fov\ to be identical for the two components.  The stellar
evolution models they used are based on the {\tt MONSTAR} code
\citep{Campbell:2008}, with physical ingredients described therein. In
particular, these models result in a mixing length parameter for the
Sun of $\amix_{\sun} = 1.60$, somewhat lower than ours (see below).

To maintain consistency with the results of Paper~II and Paper~III,
our stellar evolution tracks used to fit the binaries were calculated
here with the Modules for Experiments in Stellar Astrophysics package
\citep[MESA;][]{Paxton:2011, Paxton:2013, Paxton:2015} version 7385,
with the physical ingredients given in our previous work, and ignoring
the effects of rotation. The solar-calibrated value of the mixing
length parameter for these models and our setup is $\amix_{\sun} =
1.84$. All our calculations have included microscopic diffusion.
\cb\ did not explicitly say whether they took this effect into
account, and we are unable to determine this from an examination of
the literature sources for their stellar evolution code. We return to
this issue later (Subsection~\ref{sec:summary}). The element mixture
adopted here is that of \cite{Asplund:2009}, which is the same as used
by \cb. The corresponding mass fraction of metals for the Sun is
$Z_{\sun} = 0.0134$. The helium abundance in our model grids follows
the enrichment law $Y = 0.249 + 1.67 Z$, as in our earlier work.
CB2018 used fixed values of $Y = 0.25$ or 0.26.

In general we have chosen to explore the same ranges in \fov, \amix,
and [Fe/H] as did \cb, though in a few instances we expanded them
somewhat. Our grids for each system are uniform and complete in
\fov\ and \amix.  \cb\ did not specify the step sizes they used for
\fov\ or \amix, and from their description it does not appear that
their grids are uniformly spaced or complete.  Instead, in most cases
we presume they sampled the range manually at a few representative
values of \fov\ and \amix, as well as [Fe/H]. For the metallicity we
typically explored values within the range allowed by the measured
abundances, when available, or else we used the intervals listed by
\cb.

We note also that while three of the binaries in their Table~1 show a
range for the mixing length parameter, the other five appear to have
been treated with a single value, usually $\amix_{\sun} = 1.60$, which
is the mixing length for the Sun according to the {\tt MONSTAR} models
of \cb. Holding \amix\ fixed at a value appropriate for the Sun
carries the risk of biasing the fits for stars that are significantly
different from the Sun. Particularly in those cases, but also in
others, we have expanded the range of \amix\ values to account for the
theoretical expectation, based on 3D simulations, that \amix\ may
depend on the evolutionary state of the star or on its metallicity
\citep[see, e.g.,][]{Magic:2015}.\footnote{Due to the different input
  physics of MESA (mainly the equation of state and the opacities), a
  direct comparison of our \amix\ values with those generated by 3D
  simulations is not straightforward.}  Our ranges typically extend
toward larger values than CB2018, based on the difference seen in the
mixing length parameter for the Sun between MESA and {\tt MONSTAR}.
The larger range of \amix\ values we explore will tend to allow a
larger spread of \fov\ values. Doing this is therefore more
conservative when it comes to assigning uncertainties to the
overshooting parameter.

\cb\ considered a fit to be acceptable when the model predictions for
the two components computed at their measured masses are consistent
with the measured radii and temperatures within their reported
uncertainties. We used the same criterion here. As in our previous
work, to be conservative we accounted for possible errors in the
models themselves by allowing an age difference of up to 5\% between
the components, which again has the effect of allowing a larger range
in both \fov\ and \amix.  \cb\ did not explicitly say whether they did
something similar, or whether they required the two stars to have
precisely the same age in all cases. A further ``common sense'' rule
we have adopted concerns systems for which the observations admit
solutions with one of the components in a rapid phase of evolution,
and also solutions in a slower phase with a different \fov\ and/or
\amix. As the star can only be in one evolutionary state at a time, in
these situations we have chosen the slower phase because that scenario
is more likely a priori. \cb\ allowed both types of solutions at the
same time, which results in larger ranges of permitted \fov\ values.

As we will see below, for several systems it is also the case
that \fov\ values for the primary are better constrained than the
secondary because it is a more evolved star.  When the mass ratio is
close to unity, it is not unreasonable to expect the two stars will
also have similar degrees of overshooting, which can effectively
constrain the otherwise large range that the secondary on its own
would permit. In the next section we show how these reasonable
assumptions can be applied to some of the \cb\ systems.

\section{Results of our grid experiments}
\label{sec:grids}

We examine each of the eight DLEBs studied by \cb\ in order of
decreasing primary mass. The figures accompanying each system below
are all drawn to the same scale to permit a direct comparison.  A red
diagonal line is shown to indicate the \fov\ intervals that \cb\ claim
are allowable, which are always identical for the two components. Note
that while a first glance at their Table~1 may suggest that all
primary/secondary \fov\ combinations within their reported ranges give
acceptable fits, strictly speaking they have only explored the
diagonal of those 2D regions, as shown in the figures that follow. For
our own determinations we display the 2D regions to illustrate the
sometimes weaker constraint on \fov\ for the less evolved secondaries,
and to identify which combinations of unequal \fov\ values provide
acceptable fits.

\subsection{OGLE-LMC-ECL-CEP-0227}

\begin{figure}
\centering
\begin{tabular}{cc}
\includegraphics[width=4.1cm]{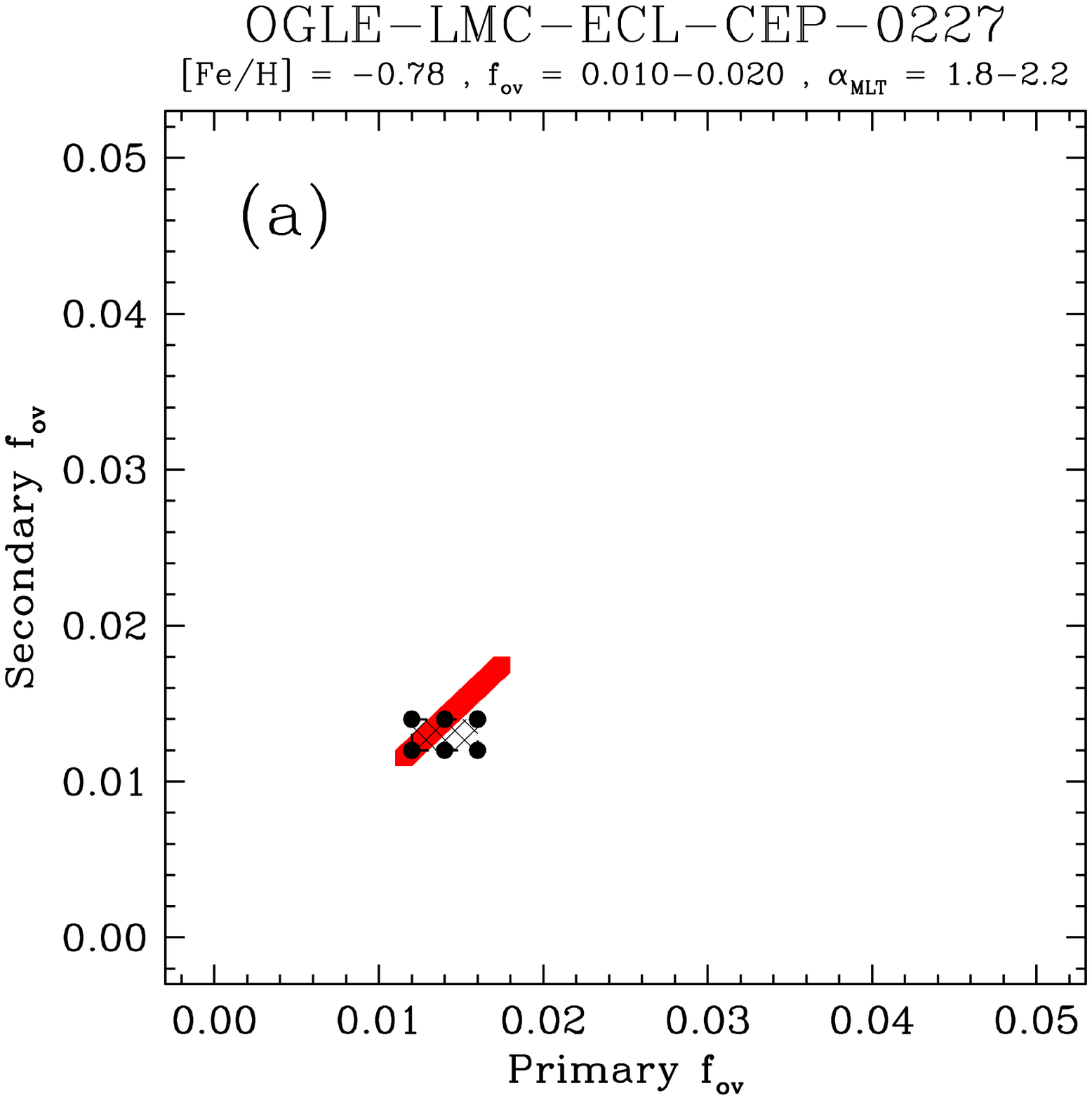} &
\includegraphics[width=4.1cm]{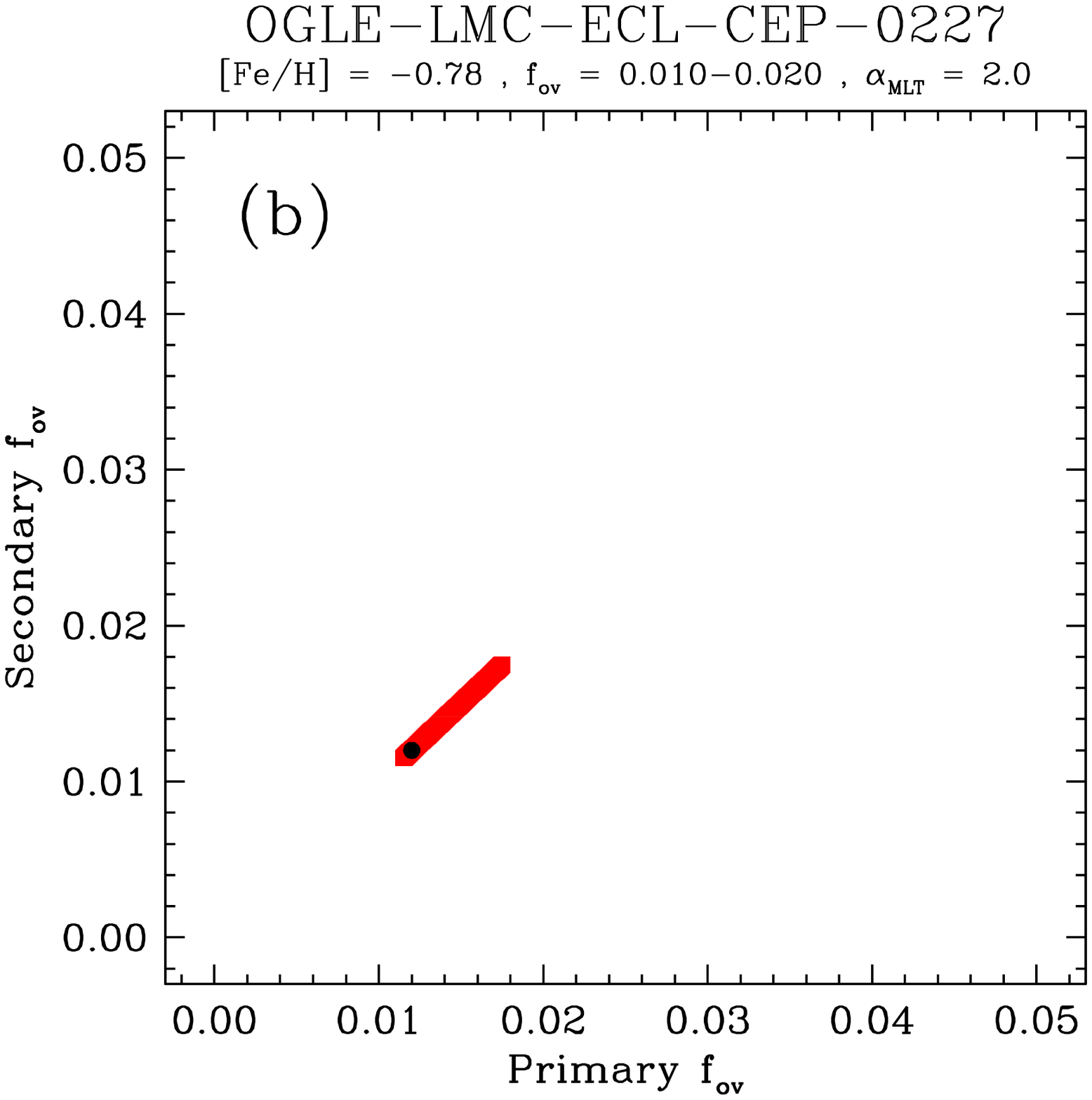}
\end{tabular}

\figcaption{(a) Values of the overshooting parameter for the primary
  and secondary of OGLE-LMC-ECL-CEP-0227 that give acceptable fits to
  the observations within our 5\% limit on the age difference
  (dots). The full ranges we find are shown by the crosshatched area.
  The red diagonal line in this and subsequent figures marks the full
  ranges in \fov\ reported by \cb. Labels indicate the adopted system
  metallicity and the range of values explored in our grids for
  \fov\ and \amix. (b) Same as panel (a), restricted to $\amix =
  2.00$. \label{fig:ogle227}}
\end{figure}

Our grid of evolutionary tracks for this system covers the \fov\ and
\amix\ intervals 0.010--0.020 and 1.80--2.20, respectively, with step
sizes of 0.002 and 0.10. Both ranges extend beyond those considered by
\cb, who used a single \amix\ value of 2.00. To our knowledge there is
no available spectroscopic metallicity for this binary. \cb\ adopted
${\rm [Fe/H]} = -1.00$, for which we find no acceptable solutions with
MESA that accomodate both component radii and both temperatures within
the observational errors, for ages differing by no more than 5\%. We
do find good fits for the metal fraction of $Z = 0.0022$ used by
\cite{Claret:2017}, corresponding to ${\rm [Fe/H]} = -0.78$ with the
\cite{Asplund:2009} element mixture, so we adopted that value
here. The acceptable fits are shown in Figure~\ref{fig:ogle227}a (dots
and crosshatched area), and restrict \fov\ more than indicated by
\cb\ despite the freedom we allowed for \amix\ and the fact that we
did not constrain the primary and secondary values of \fov\ to be
identical. As we cannot make a direct comparison using the same
metallicity they adopted, we have at least made it for the same
\amix\ value they chose (2.00). Under these conditions we find a
single satisfactory solution, indicated in the right panel.

\setlength{\tabcolsep}{2pt} 
\setlength{\tabcolsep}{6pt} 

\subsection{LMC-562.05$-$9009}

\setlength{\tabcolsep}{2pt} 
\begin{figure}
\centering
\begin{tabular}{cc}
\includegraphics[width=4.1cm]{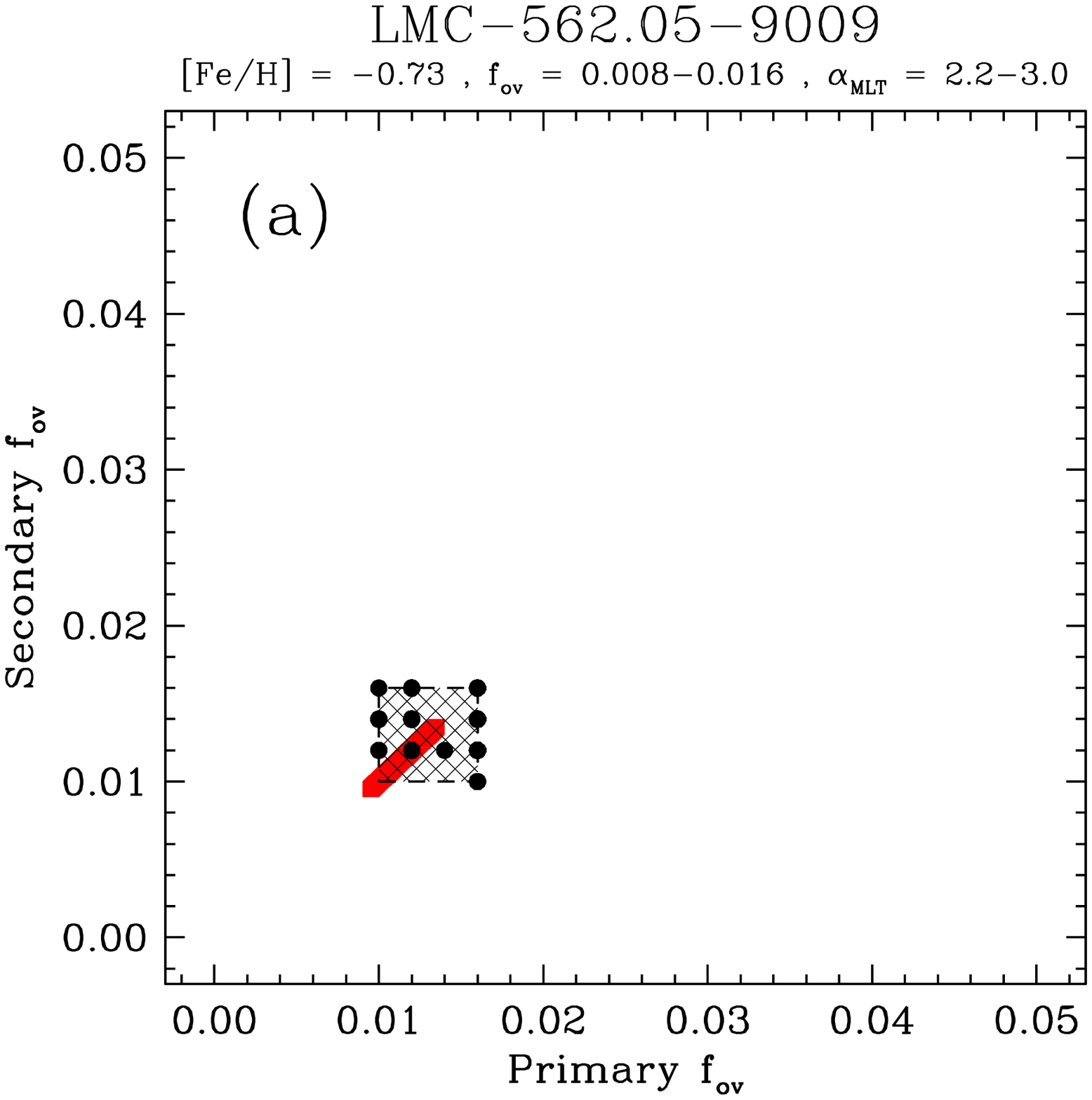} &
\includegraphics[width=4.1cm]{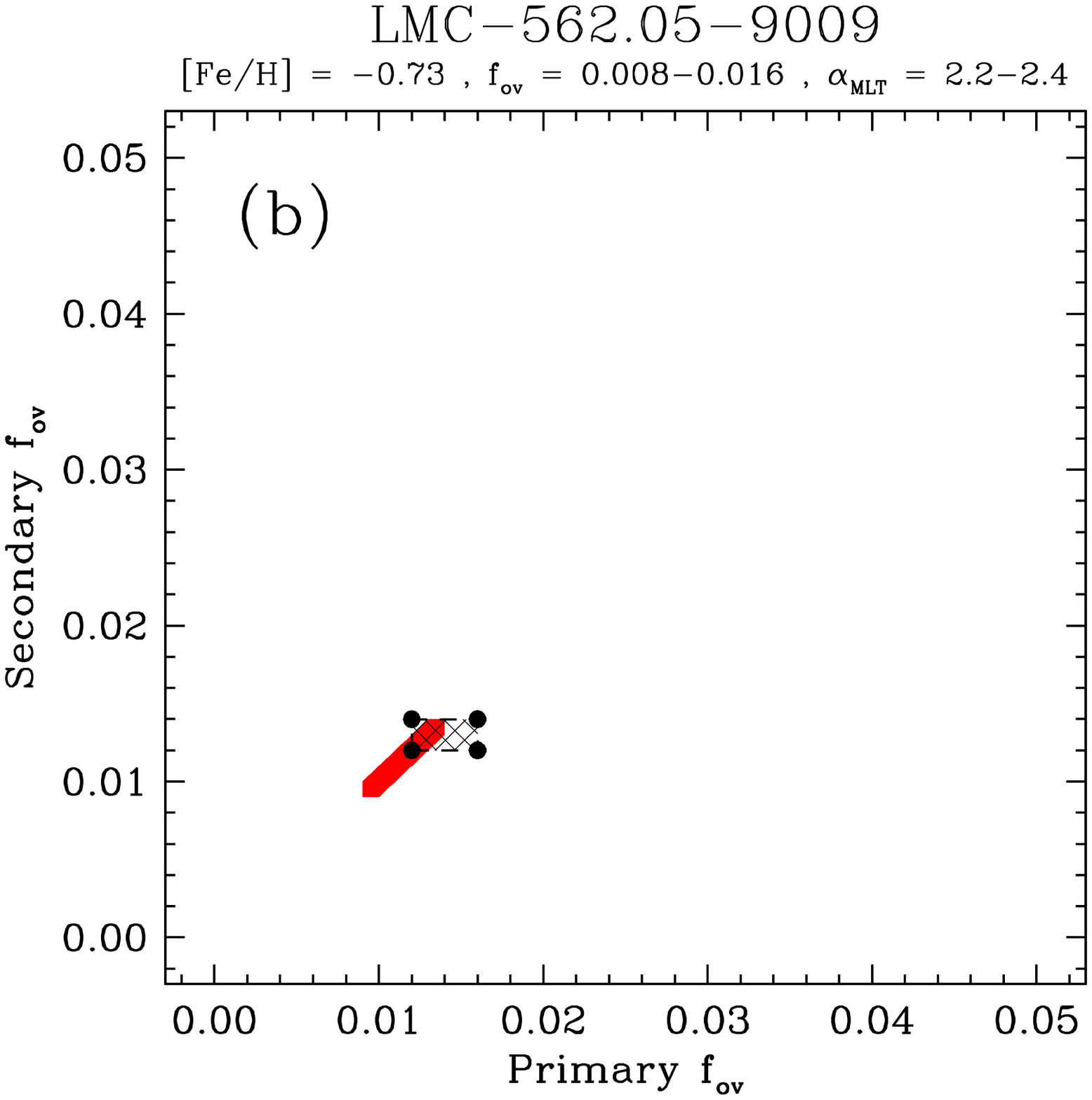}
\end{tabular}
\figcaption{(a) Same as Figure~\ref{fig:ogle227} for
  LMC-562.05$-$9009. (b) \amix\ values restricted to the
  interval 2.20--2.40 (see text).\label{fig:ogle562}}
\end{figure}
\setlength{\tabcolsep}{6pt} 

For this binary our grids were computed over the \fov\ range of
0.008--0.016, with a step size of 0.002, and a range in \amix\ of
2.20--3.00, in intervals of 0.10. Again these are both slightly larger
than those of \cb. This is another system lacking a spectroscopic
metallicity measurement, so we adopted a value of ${\rm [Fe/H]} =
-0.73$ ($Z = 0.0025$), close to the \cb\ choice. The solutions
reported by \cb\ for this binary are quite similar to ours, and are
shown in Figure~\ref{fig:ogle562}a. One significant difference,
however, is that they allowed values of \amix\ as high as 3.00, which
exceeds the maximum values predicted by the models of
\cite{Magic:2015}, based on full 3D radiative hydrodynamic
calculations. In the right panel of the figure we have restricted
\amix\ to more moderate values between 2.20 and 2.40, which results in
a smaller range of allowed values of \fov\ than reported by \cb.

\subsection{$\chi^2$ Hya}

\begin{figure}
\centering
\begin{tabular}{cc}
\includegraphics[width=4.1cm]{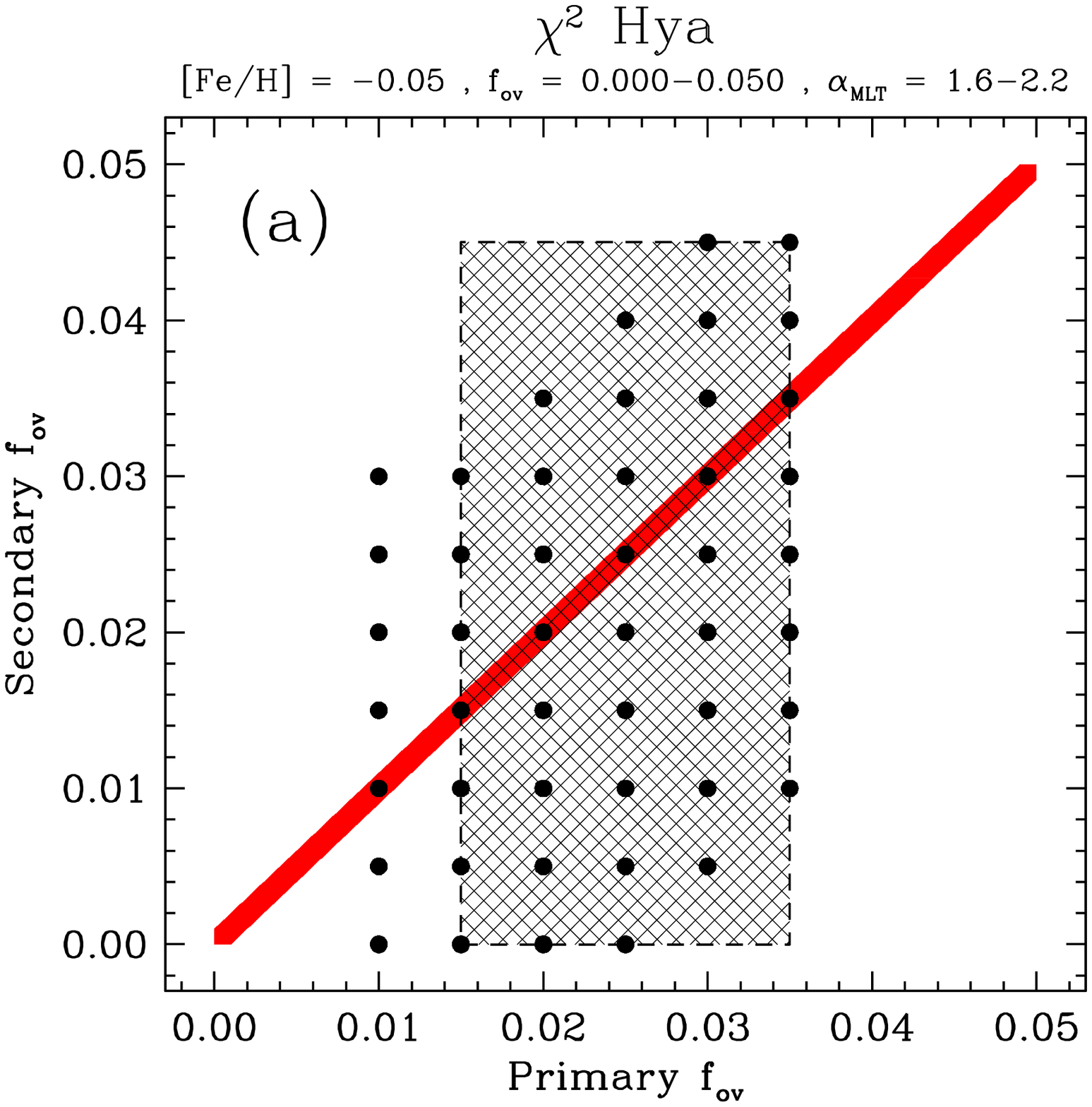} &
\includegraphics[width=4.1cm]{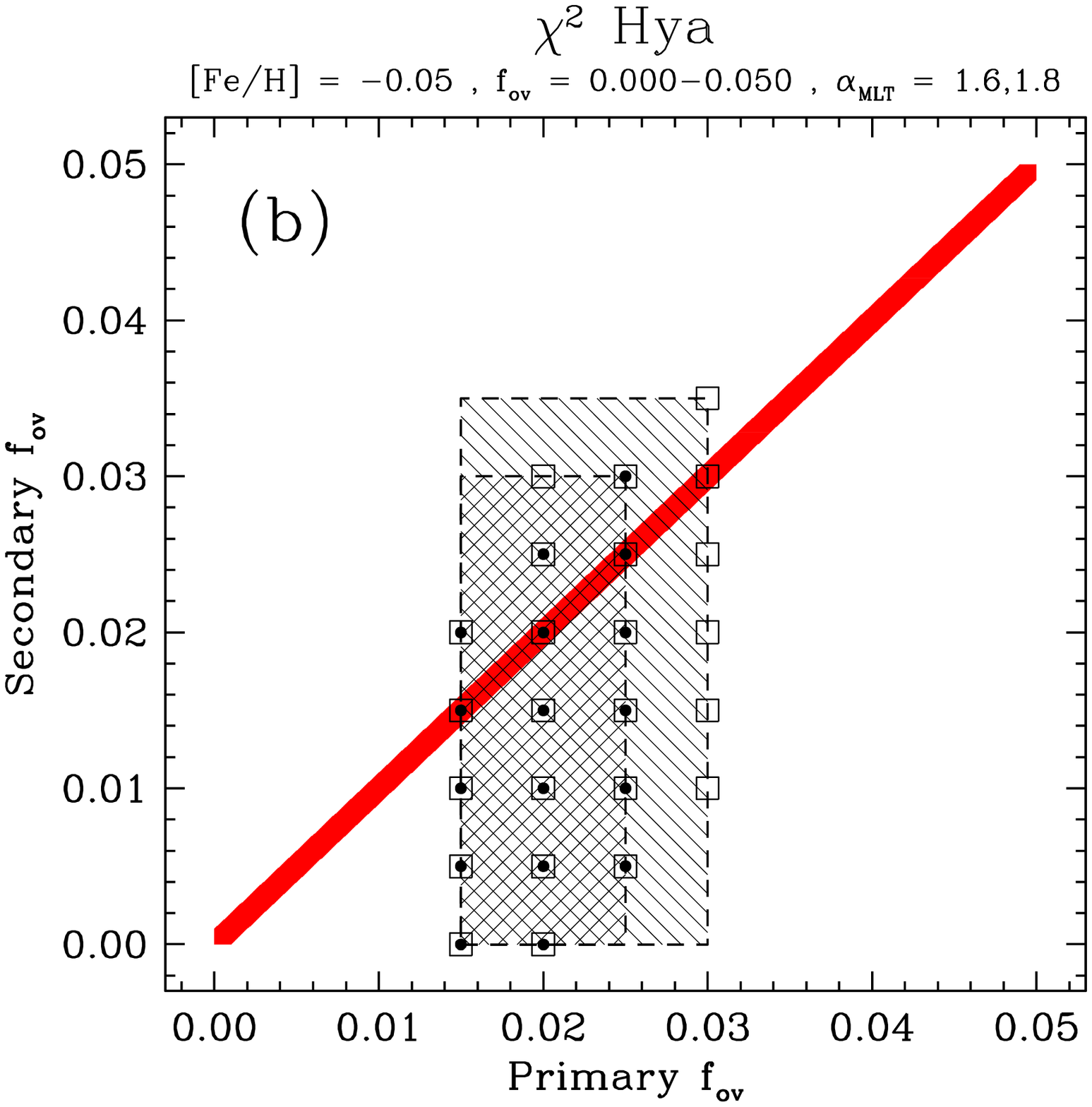} \\ [1ex]
\includegraphics[width=4.1cm]{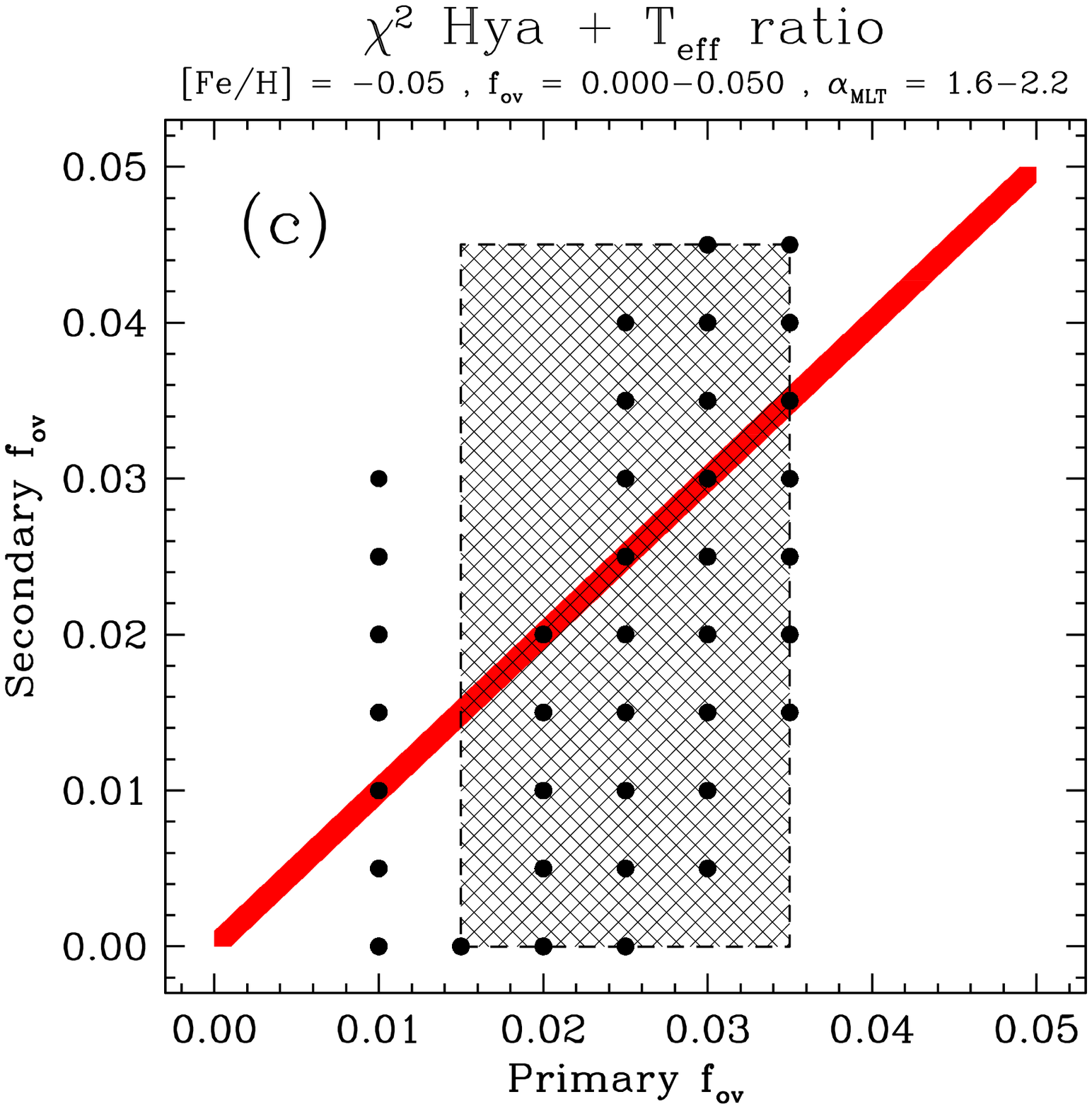} &
\includegraphics[width=4.1cm]{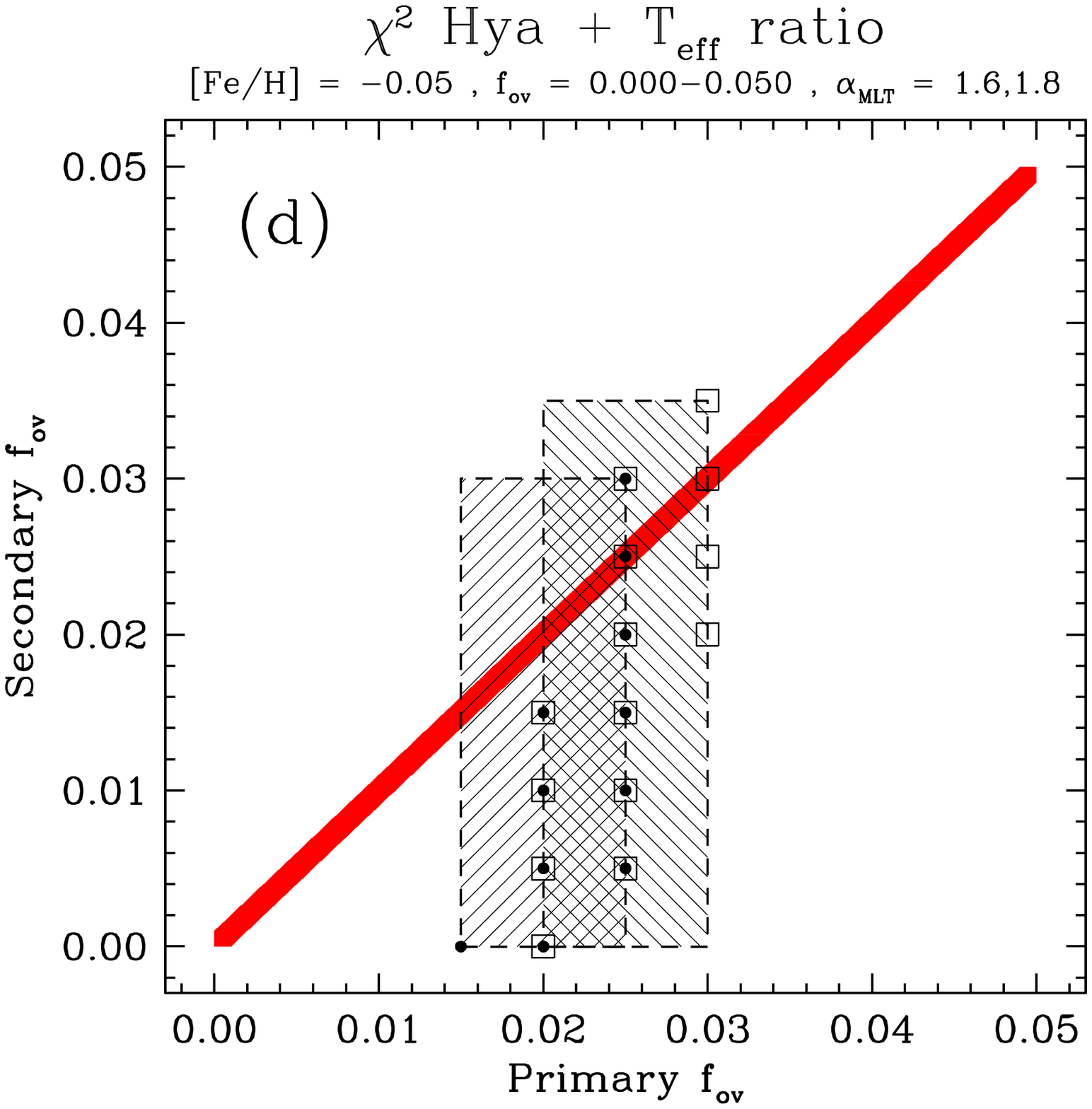}
\end{tabular}

\figcaption{(a) Same as Figure~\ref{fig:ogle227} for $\chi^2$~Hya.
  Primary values of $\fov = 0.010$ are in the rapid phase of evolution
  and are regarded as less likely. (b) \amix\ for both stars
  restricted to the \cb\ value of $\amix_{\sun} = 1.60$ (circles), or
  to a value of 1.80 near our solar-calibrated mixing length
  (squares). The corresponding viable areas of parameter space are
  indicated by the hatched regions. (c) Same as panel (a), adding the
  constraint from the measured temperature ratio (see text). (d) Same
  as panel (b), adding the constraint from the measured temperature
  ratio. \label{fig:chi2hya}}
\end{figure}

This particularly interesting binary system features rather unequal
masses \citep[$3.605 \pm 0.078~M_{\sun}$ and $2.632 \pm
  0.049~M_{\sun}$;][]{Torres:2010}. \cb\ used the same \fov\ values
for the primary and secondary. Our models for each star were computed
over the same \fov\ range as theirs (0.000--0.050) in steps of 0.005,
and the \amix\ values we explored are between 1.60 and 2.20 with our
usual step size of 0.10. \cb\ considered a single mixing length
parameter of 1.60 for both components (the {\tt MONSTAR}
$\amix_{\sun}$). No metallicity determination is available for
$\chi^2$~Hya, so we investigated the [Fe/H] range explored by
\cb\ between $-0.15$ and the solar composition. The MESA models give
no acceptable solutions at either extreme, but we do find good fits
for an intermediate value ${\rm [Fe/H]} = -0.05$, shown in
Figure~\ref{fig:chi2hya}a. At primary \fov\ values of 0.035 the model
predictions are barely within observational errors for that star, and
only for mixing length parameters $\amix \ge 2.00$, which are
considerably higher than the value adopted by \cb. For $\fov \le
0.010$ the primary is in a rapid evolutionary phase; we disregard
those solutions in favor of those at slower, more likely phases that
are permitted at higher values of \fov. The full range of values that
the observations allow for the primary and secondary is indicated by
the crosshatched region.

\setlength{\tabcolsep}{2pt} 
\setlength{\tabcolsep}{6pt} 

If we consider only the single mixing length value adopted by
\cb\ (1.60), the acceptable solutions span smaller ranges in \fov\ for
both components (Figure~\ref{fig:chi2hya}b), indicating stronger
constraints on overshooting than reported by \cb\ under similar
conditions. However, given that our solar value of \amix\ is
considerably larger than theirs, for a more fair comparison with
\cb\ we show in this latter figure the good fits for $\amix = 1.80$
(square symbols), which is the closest in our grid to the MESA
solar-calibrated value of $\amix_{\sun} = 1.84$. We find slightly more
extended ranges in \fov\ for both stars than with $\amix = 1.60$, but
still appreciably smaller than indicated by \cb.

$\chi^2$~Hya is special in that it has a well measured temperature
ratio of $T_{\rm eff,2}/T_{\rm eff,1} = 0.945 \pm 0.009$ from the work
of \cite{Clausen:1978}, which sets a strong additional constraint.
This quantity is closely related to the difference in the eclipse
depths, and is usually much more accurate than the individual absolute
temperatures. We find that adding the requirement that the predicted
temperature ratio be within the observational error restricts
\fov\ even further. This is shown in Figure~\ref{fig:chi2hya}c for
grids over the same \amix\ range as before, and in
Figure~\ref{fig:chi2hya}d for fixed \amix\ values of 1.60 and 1.80.

Regarding the secondary component of $\chi^2$~Hya, it is worth
pointing out that even though models formally allow them, overshooting
parameters as low as zero or 0.005 are probably unrealistic for a star
of this mass ($M = 2.632~M_{\sun}$).  To our knowledge there is no
empirical evidence of such small values of \fov\ for similar stars
\citep[see, e.g.,][]{Schroder:1997, Stancliffe:2015, Torres:2015,
  Moravveji:2015, Moravveji:2016, Valle:2017}. With this
consideration, the uncertainty in \fov\ would be reduced even further.

\subsection{OGLE-LMC-ECL-26122}

The stellar parameters adopted by \cb\ for this binary are the
  same as we used in Paper~II, based on the work of
  \cite{Pietrzynski:2013}. Slight improvements have recently been
  reported by \cite{Graczyk:2018}, who also changed the designation
  to OGLE-LMC-SC9-230659. Here we continue to use the original values
  for a proper comparison with \cb, as well as the original
  designation to avoid confusion. An updated analysis is presented
  below in Section~\ref{sec:newbinaries}.

\begin{figure}[t!]
\epsscale{0.55}
\plotone{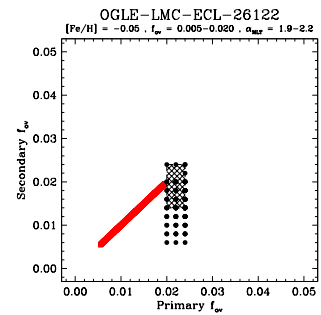}

\figcaption{Same as Figure~\ref{fig:ogle227} for
  OGLE-LMC-ECL-26122. The crosshatched area indicates favored
  solutions with the secondary in a slow phase of
  evolution. \label{fig:ogle261}}
\end{figure}

For this system our grids cover a broader interval in \fov\ than
explored by \cb\ (0.005--0.024, step size = 0.002), and the same range
in \amix\ (1.90--2.20, step size = 0.10). \cite{Pietrzynski:2013} 
reported a spectroscopic metallicity estimate for the binary of ${\rm
  [Fe/H]} = -0.15 \pm 0.10$. It is worth noting, however, that
\cb\ chose to use considerably lower values reaching ${\rm [Fe/H]} =
-0.50$.\footnote{\cb\ stated that this value ``is only marginally more
  metal-poor than the best fit from \cite{Claret:2017}, $Z =
  0.0050$''. The value we reported in Paper~II is actually $Z =
  0.0070$, corresponding to ${\rm [Fe/H]} = -0.28$, which is much
  closer to the lower limit of the measured value.  It is possible
  that CB2018 mistakenly quoted the $Z$ value from our earlier Paper~I
  (0.0050), which is based on a different element mixture. That lower
  value was found in Paper~II to be biased too low due to the use of
  an outdated value for the primordial helium abundance.}  Restricting
ourselves to the measured range only, we find no satisfactory
solutions at the upper limit of ${\rm [Fe/H]} = -0.05$, but do find
them at the lower limit, ${\rm [Fe/H]} = -0.25$. These fits are shown
in Figure~\ref{fig:ogle261} with filled circles, and indicate the
observations provide very strong constraints on the primary value of
\fov, which can only run between 0.020 to 0.024. For the secondary
component, values smaller than about 0.014 place it in a rapid phase
of evolution that we consider less likely than allowable slower phases
corresponding to a higher degree of overshooting. These more likely
fits are indicated by the crosshatched area in the figure.  Once again
the relatively similar masses of the components \citep[$3.593 \pm
  0.055~M_{\sun}$ and $3.411 \pm 0.047~M_{\sun}$;][]{Pietrzynski:2013}
would enable us to restrict the secondary interval even further by
requiring its \fov\ to be similar to that of the primary, as
\cb\ assumed. In conclusion, we find that the range of \fov\ values
permitted by the observations of OGLE-LMC-ECL-26122 is considerably
smaller than proposed by \cb.

\subsection{SZ Cen}

\begin{figure}[t!]
\centering
\begin{tabular}{cc}
\includegraphics[width=4.1cm]{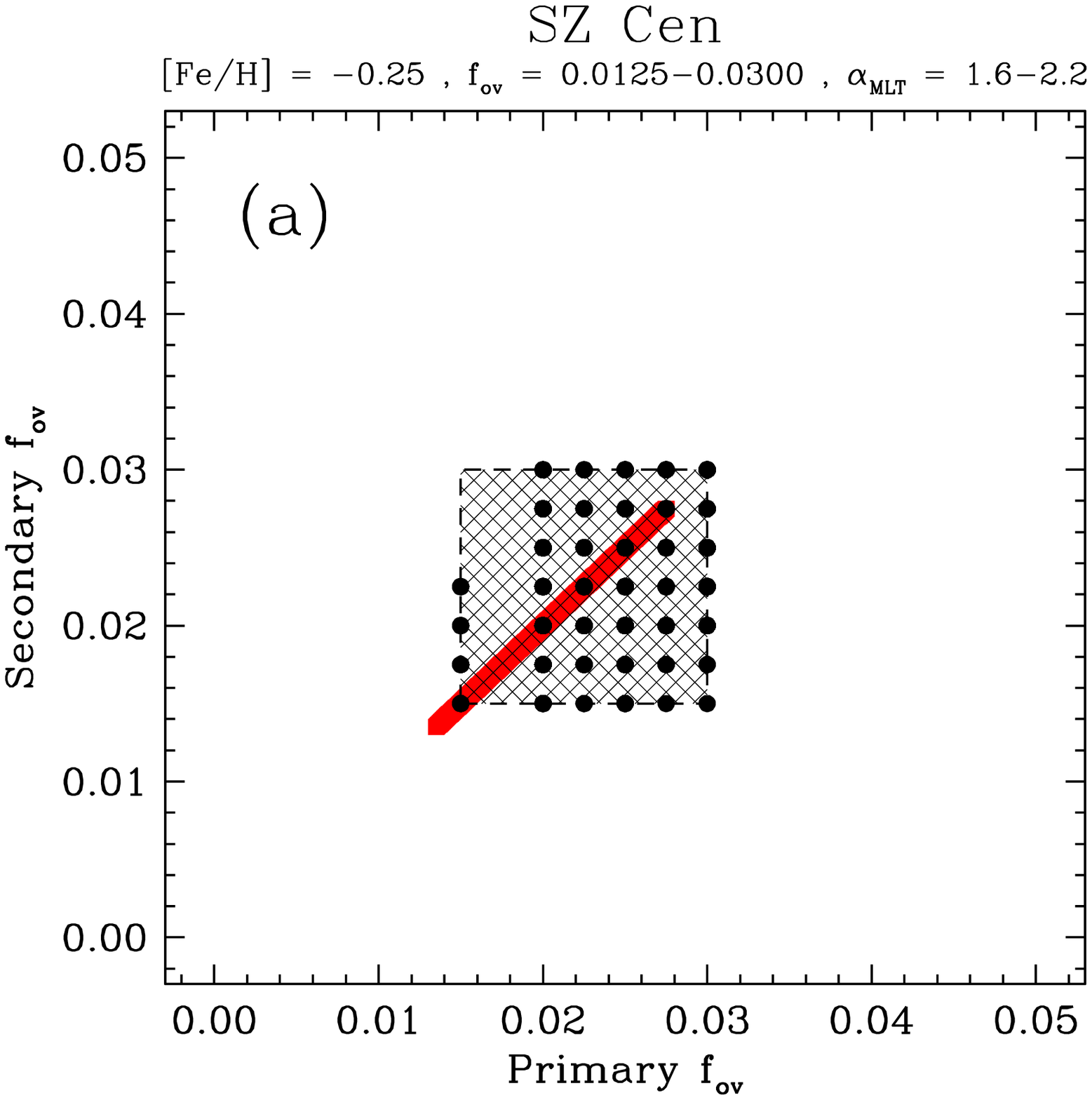} &
\includegraphics[width=4.1cm]{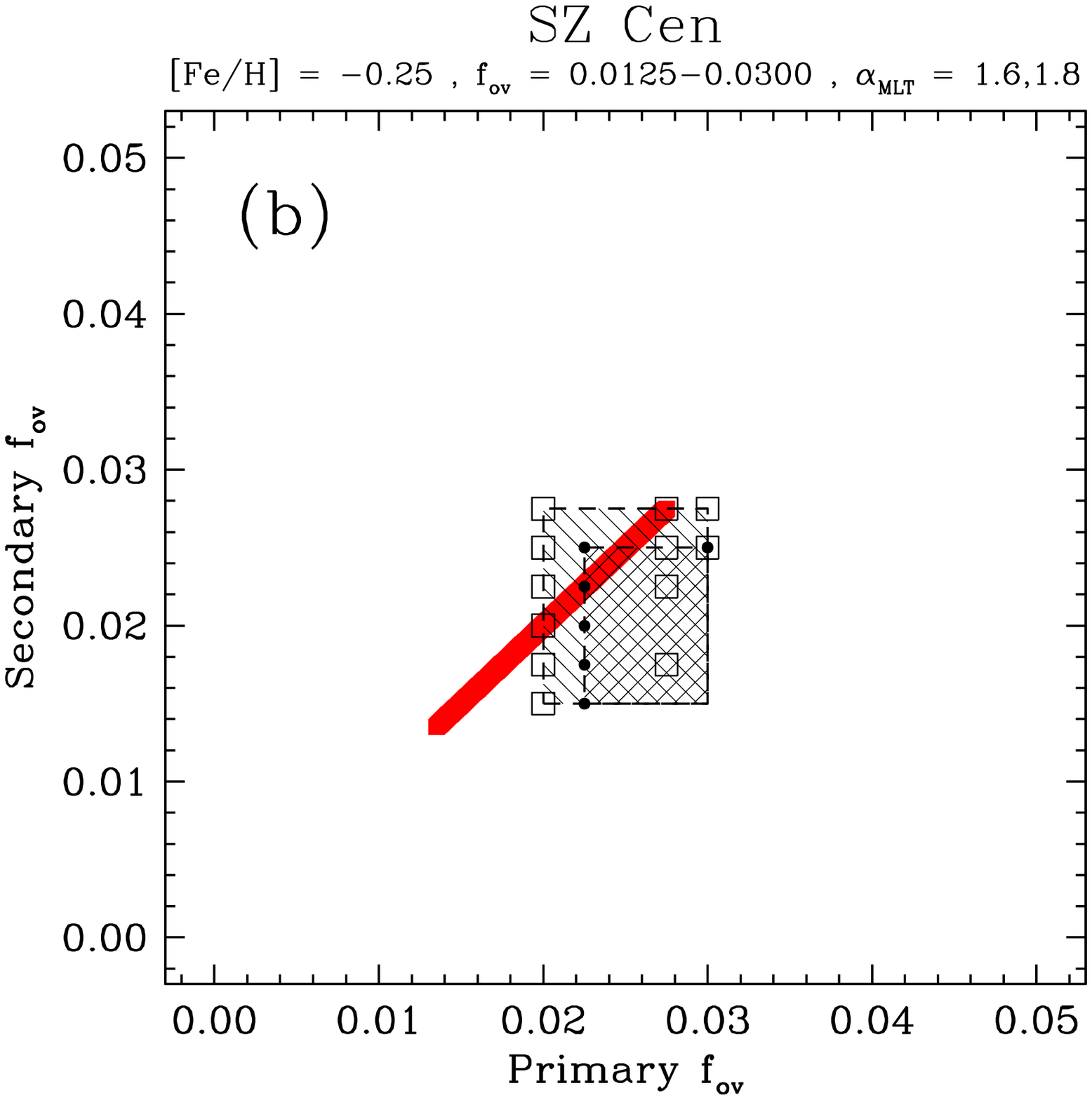} \\ [1ex]
\includegraphics[width=4.1cm]{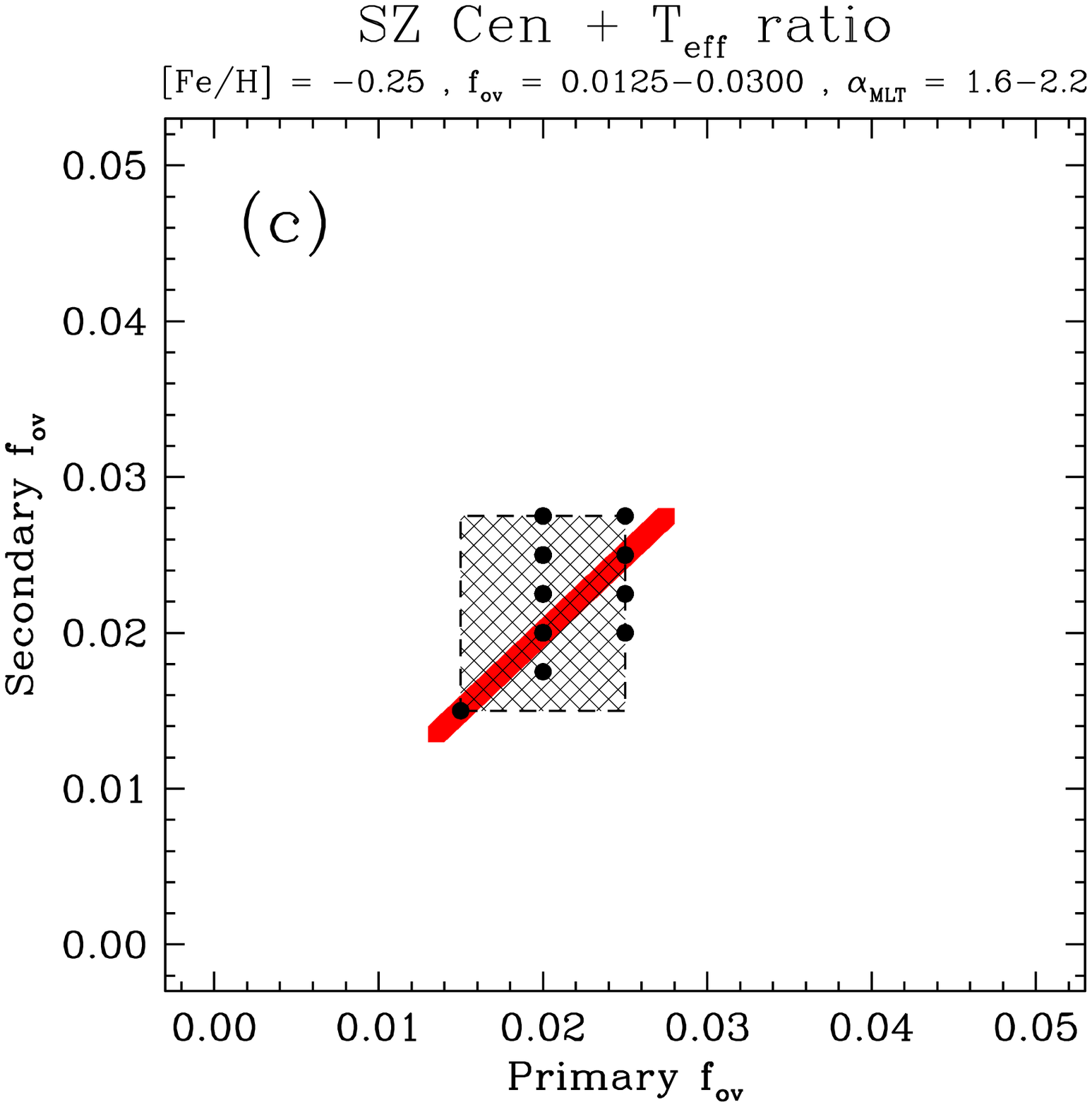} &
\includegraphics[width=4.1cm]{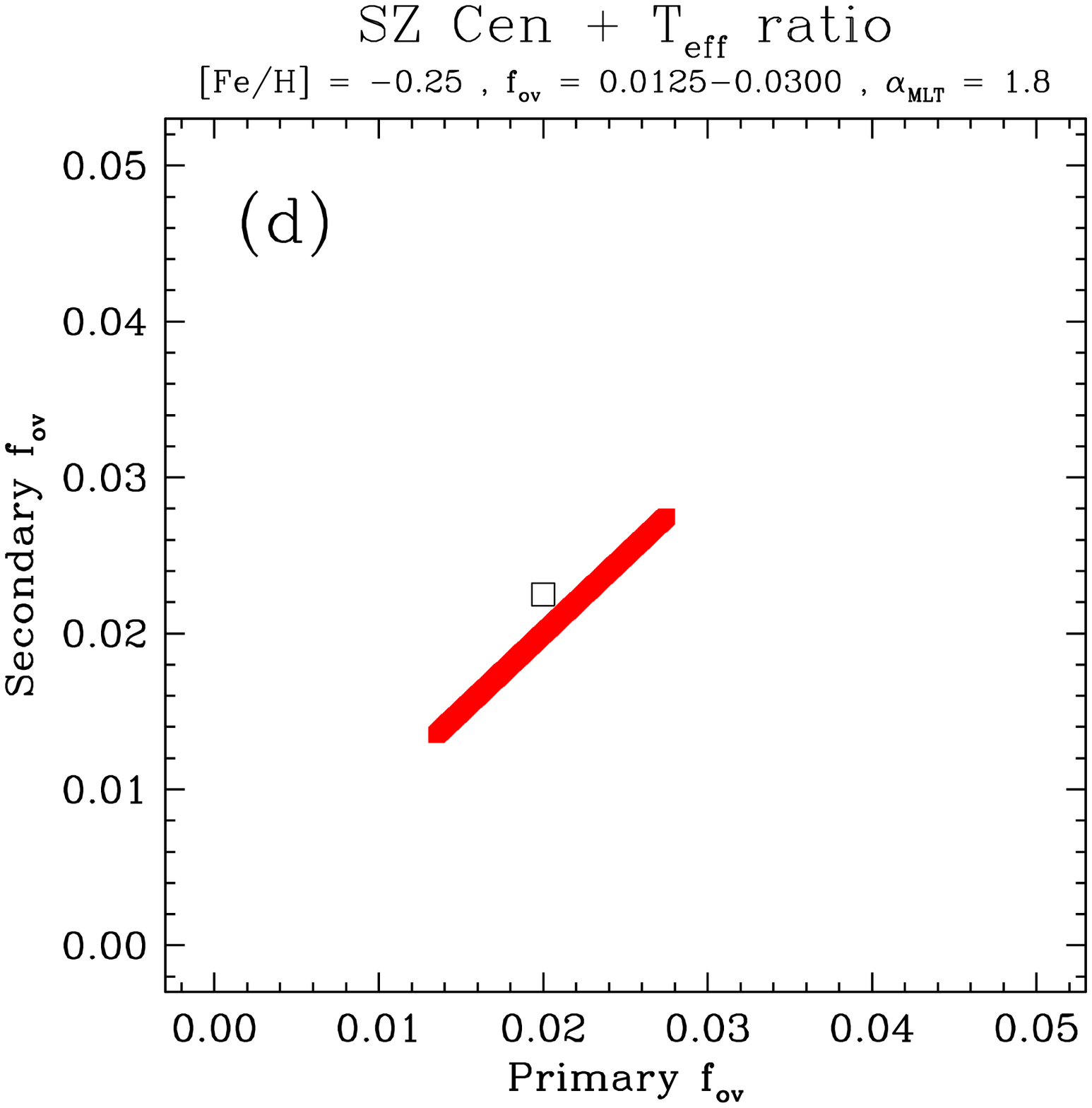} \\ [2ex]
\includegraphics[width=4.1cm]{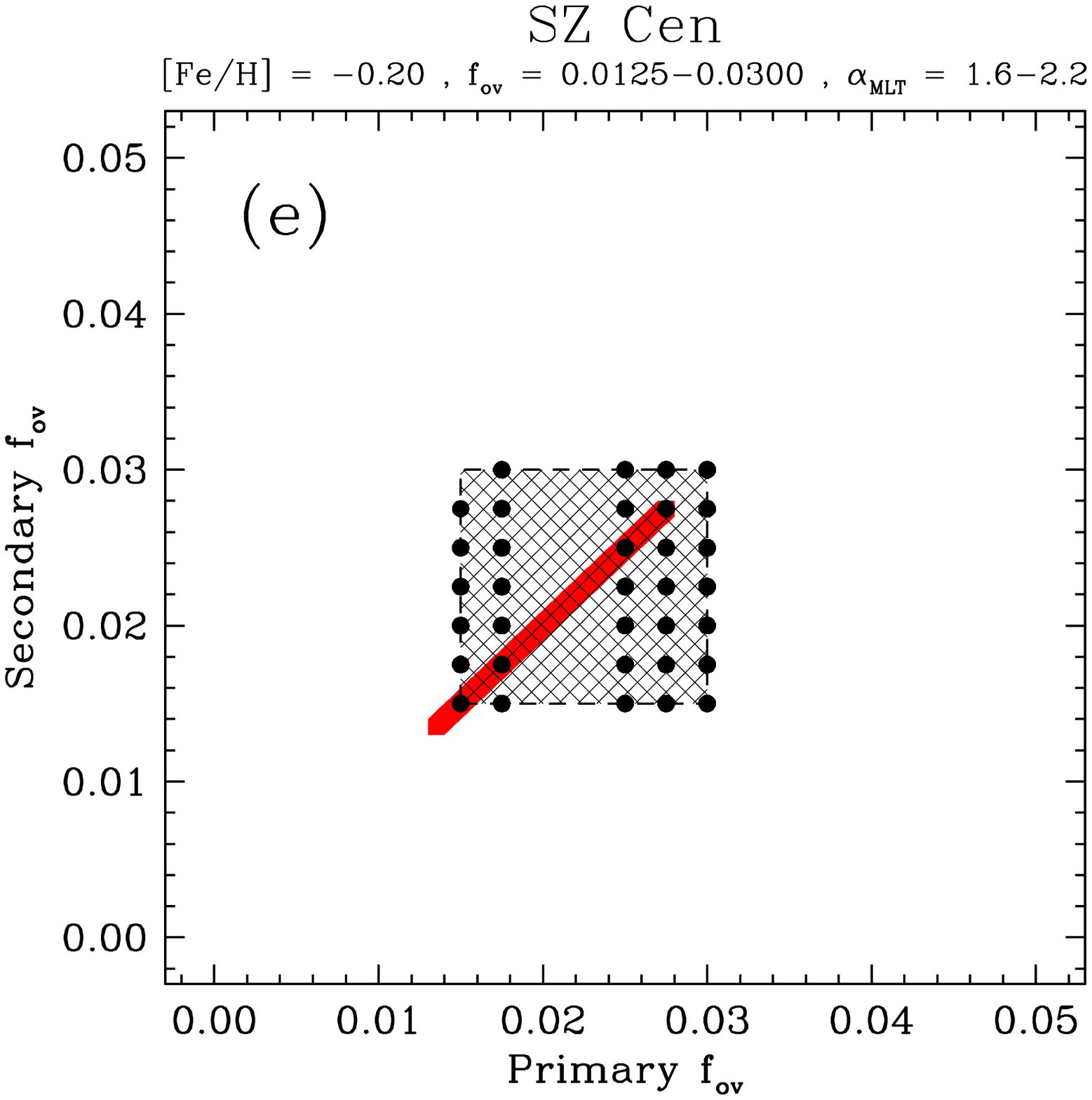} &
\includegraphics[width=4.1cm]{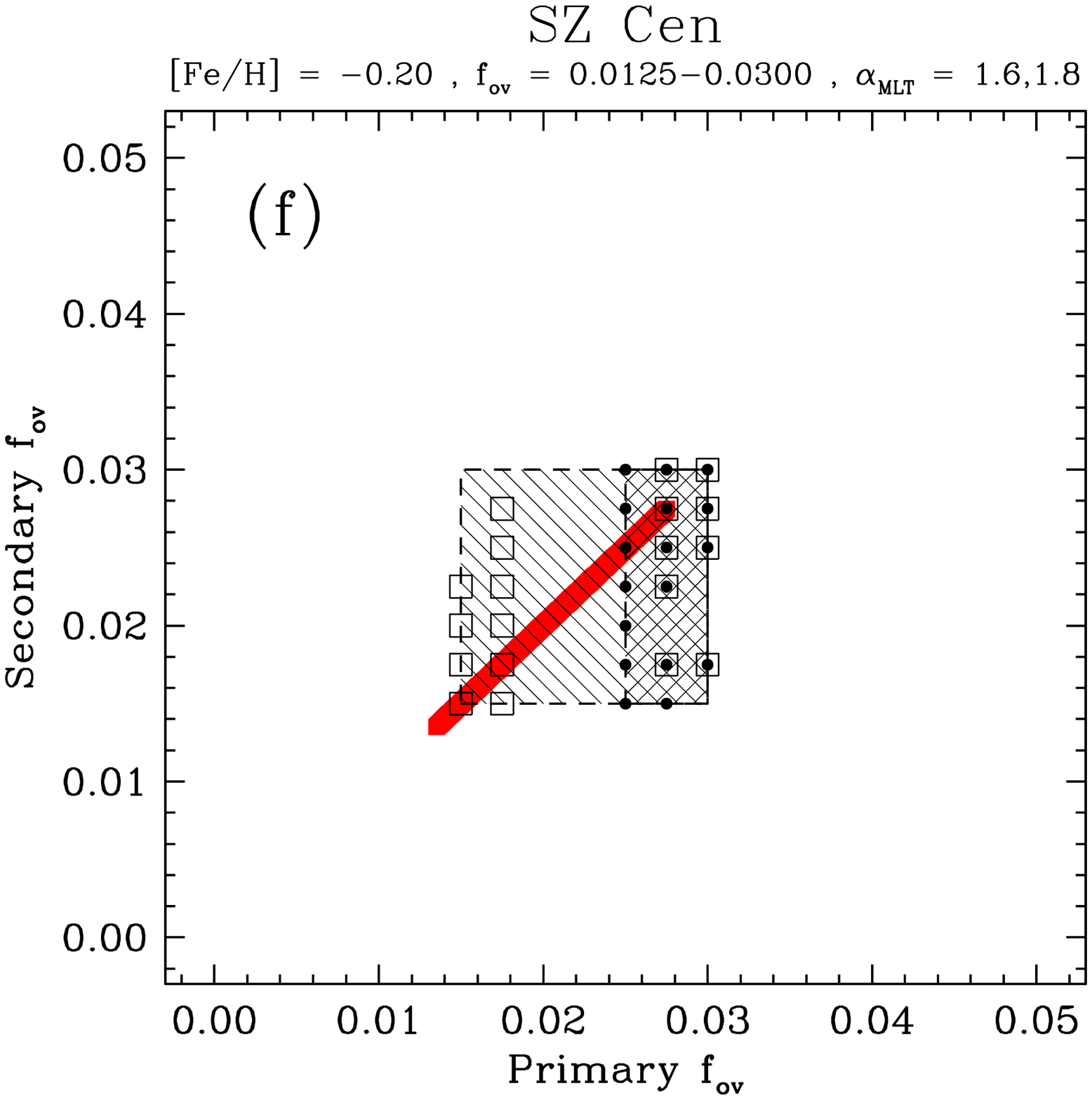} \\ [1ex]
\includegraphics[width=4.1cm]{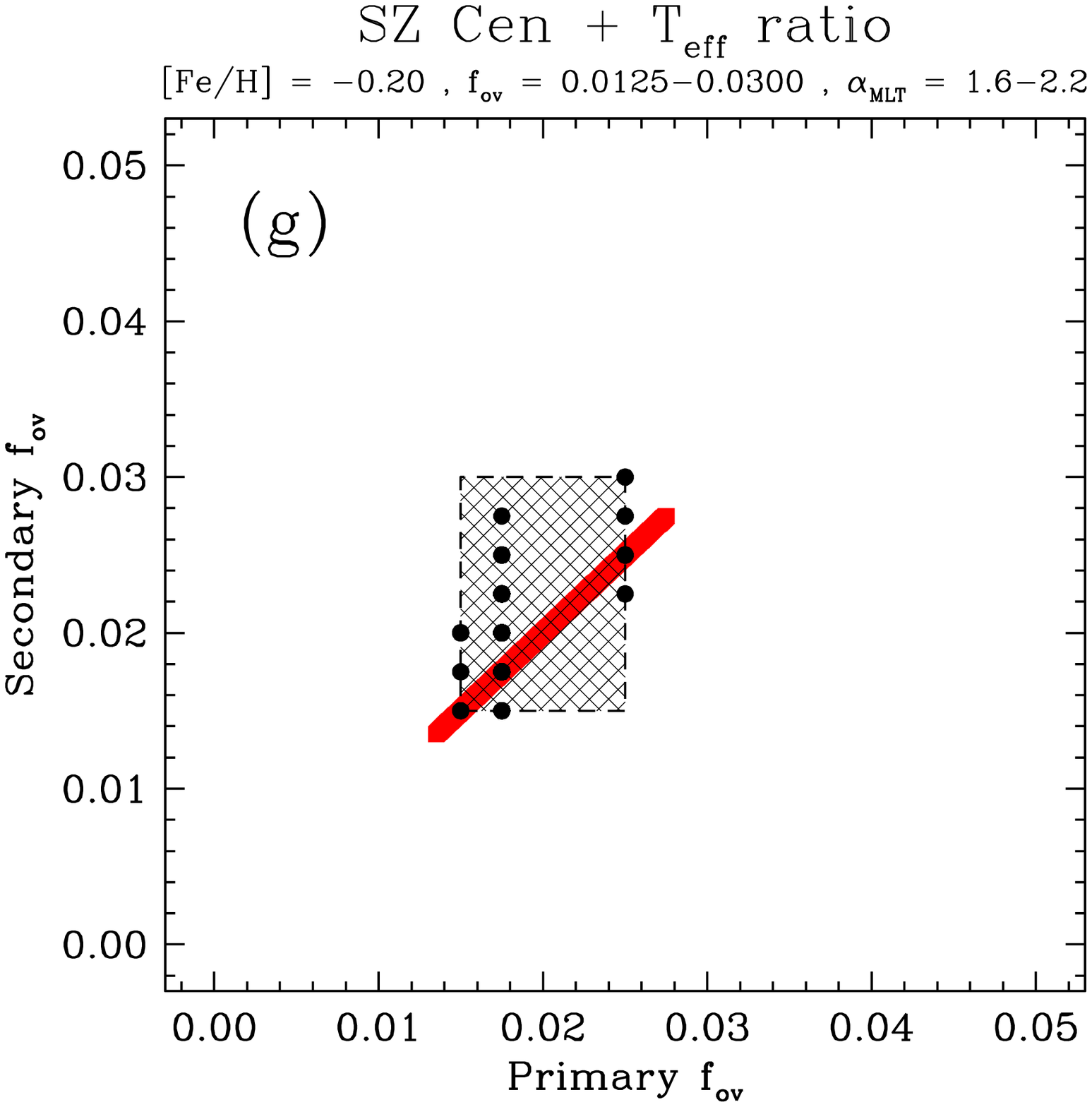} &
\includegraphics[width=4.1cm]{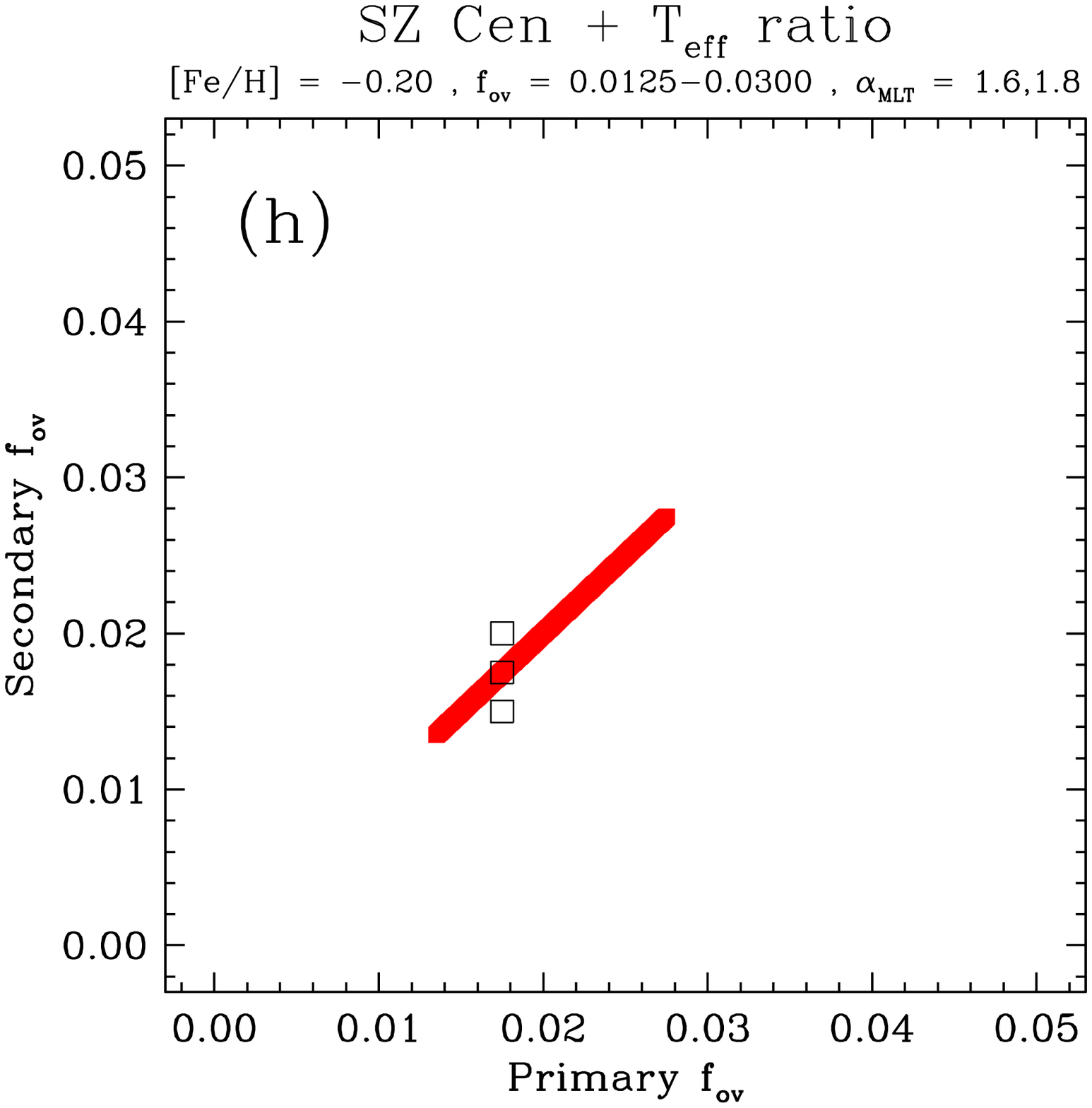}
\end{tabular}
\figcaption{(a) Same as Figure~\ref{fig:ogle227} for SZ~Cen, with
  ${\rm [Fe/H]} = -0.25$. (b) \amix\ for both stars restricted to the
  \cb\ value of $\amix_{\sun} = 1.60$ (circles), or to a value of
  1.80 near our solar-calibrated mixing length (squares). (c) Same as
  panel (a), adding the constraint from the measured temperature ratio
  (see text). (d) Same as panel (b), adding the constraint from the
  measured temperature ratio. (e)-(h) Same as panels (a)-(d), for
  ${\rm [Fe/H]} = -0.20$. \label{fig:szcen}}
\end{figure}

The grids were computed for \fov\ values between 0.0125 and 0.0300 in
steps of 0.0025, and \amix\ values between 1.60 and 2.20 every 0.10.
\cb\ used only their solar-calibrated mixing length parameter of
$\amix_{\sun} = 1.60$, and considered a very narrow metallicity range
between ${\rm [Fe/H]} = -0.25$ and $-0.20$. There is no measured
metallicity available for this binary. The acceptable solutions we
find for ${\rm [Fe/H]} = -0.25$ are shown in Figure~\ref{fig:szcen}a,
where we include those with $\fov = 0.030$ for the secondary even
though they barely satisfy the observations within observational
errors (and only for \amix\ well above 1.60).  Selecting only the fits
with $\amix = 1.60$, to match what \cb\ did, results in a reduced
range of permitted \fov\ values (Figure~\ref{fig:szcen}b, circles),
which is smaller than reported by \cb.  Solutions with $\amix = 1.80$,
close to our own solar-calibrated value, are shown with
squares. Hatched regions cover the full range for the primary and
secondary in each case.

\setlength{\tabcolsep}{2pt} 
\setlength{\tabcolsep}{6pt} 

SZ~Cen is another system with a highly precise measurement of the
temperature ratio: $T_{\rm eff,2}/T_{\rm eff,1} = 1.035 \pm 0.003$
\citep{Gronbech:1977}. If we apply this additional observational
constraint under the same conditions as before, we obtain the viable
solutions shown in Figures~\ref{fig:szcen}c (full range of \amix) and
\ref{fig:szcen}d (restricted \amix). In the latter panel there are no
acceptable fits with $\amix = 1.60$, and only a single one with $\amix
= 1.80$, showing how much better \fov\ can be determined.

The results for the other end of the metallicity range, ${\rm [Fe/H]}
= -0.20$, are shown in Figure~\ref{fig:szcen}e-\ref{fig:szcen}h using
the same parameters as in
Figure~\ref{fig:szcen}a-\ref{fig:szcen}d. The allowed \fov\ ranges are
quite similar for the two compositions.

\subsection{AY Cam}

\begin{figure}
\centering
\begin{tabular}{cc}
\includegraphics[width=4.1cm]{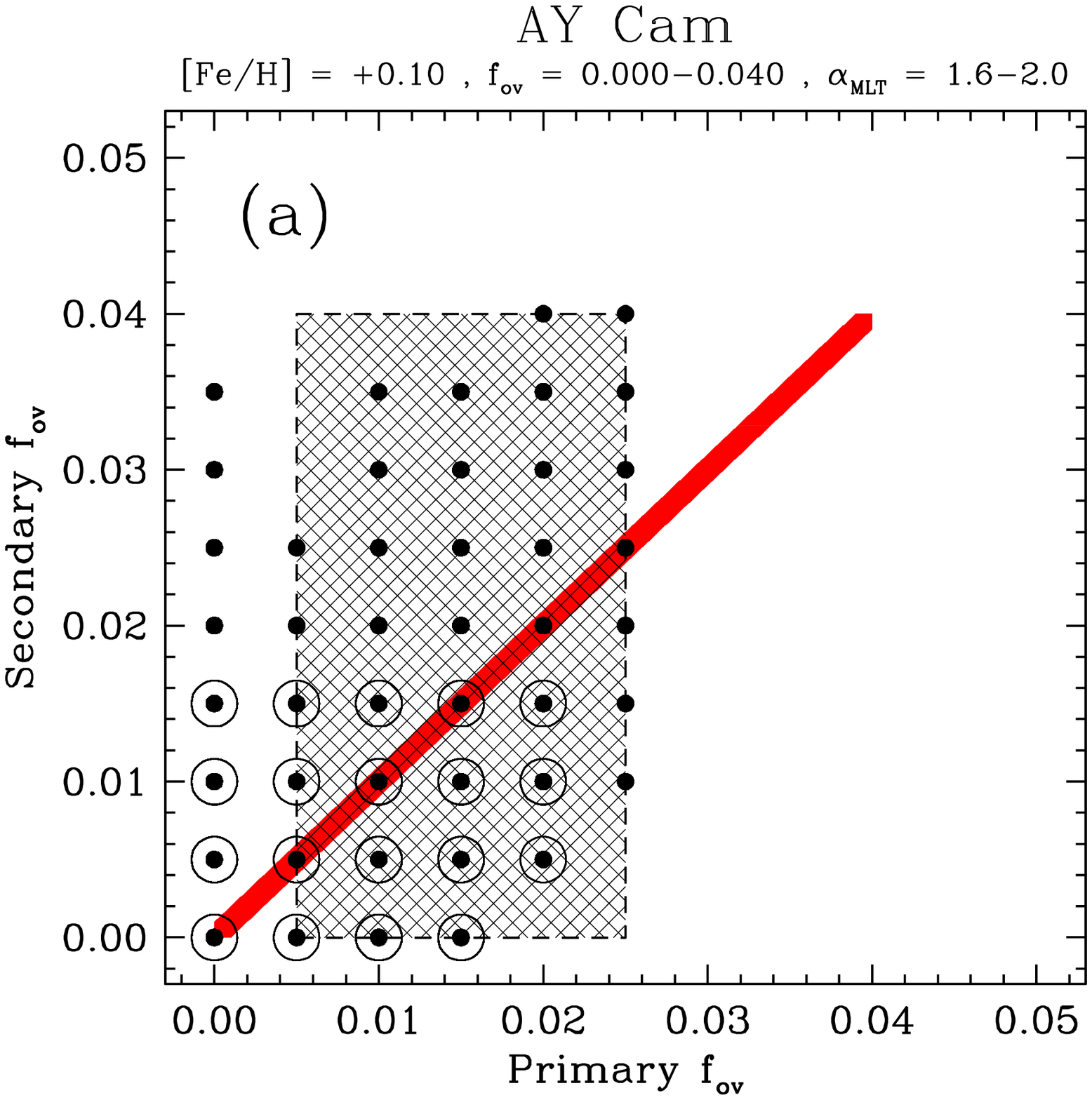} &
\includegraphics[width=4.1cm]{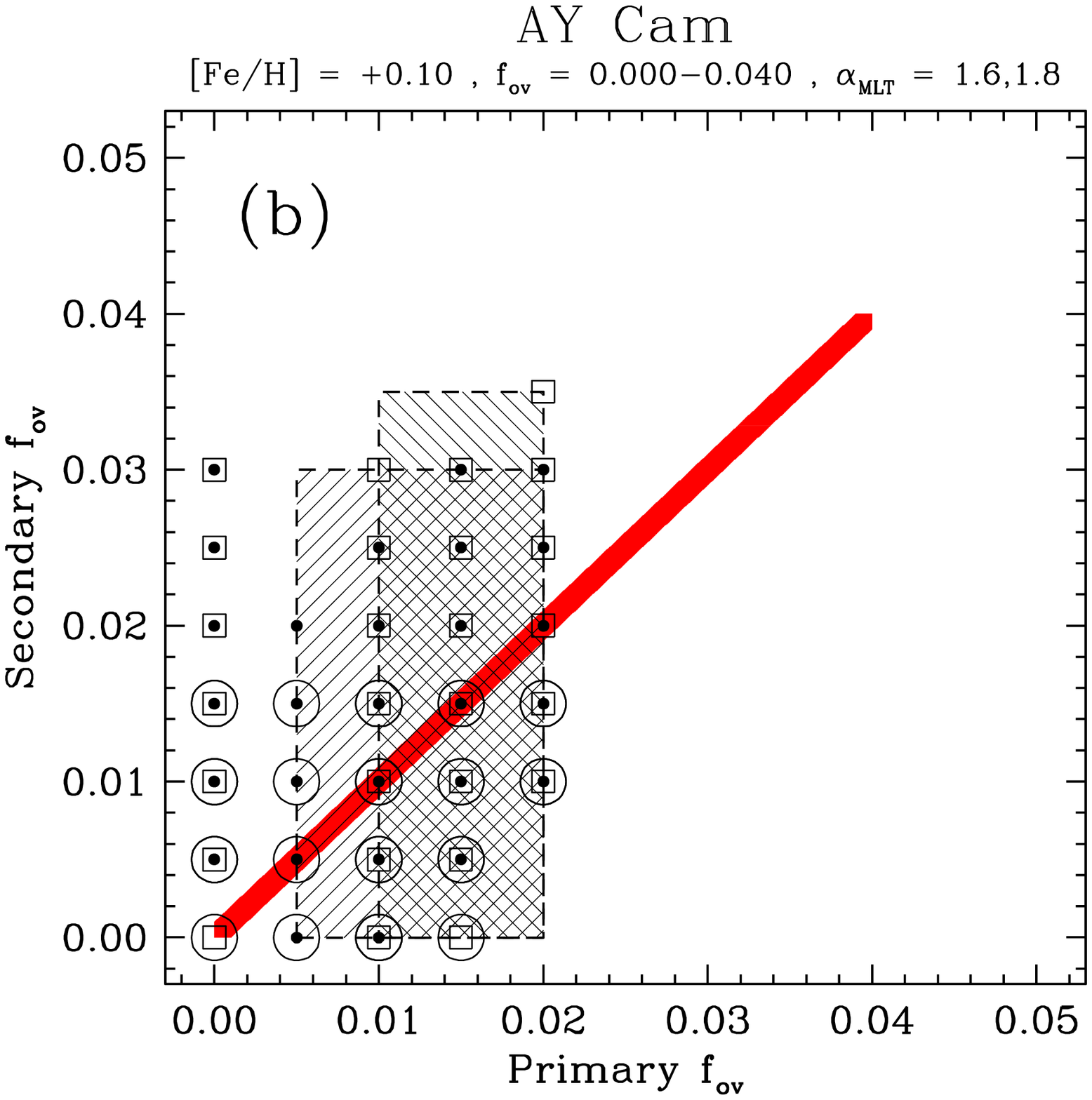}
\end{tabular}
\figcaption{(a) Same as Figure~\ref{fig:ogle227} for AY~Cam. (b)
  Acceptable fits for \amix\ set to the \cb\ value $\amix_{\sun} =
  1.60$ (circles), or to a value of 1.80 near our solar-calibrated
  mixing length (squares). Open circles in both panels mark solutions
  with \fov\ within the upper limit established by
  \cite{Roxburgh:1992} for stars of this mass (see
  Appendix~\ref{sec:theory}). \label{fig:aycam}}
\end{figure}

Our grids of stellar evolution tracks span the full interval of
\fov\ values considered by \cb\ (0.000--0.040), and were calculated
with a step of 0.005. Our range in \amix\ is 1.60--2.00 in steps of
0.10, whereas \cb\ used only the fixed value $\amix_{\sun} = 1.60$.
AY~Cam lacks a spectroscopic metallicity estimate, and the [Fe/H]
range considered by \cb\ is fairly small. We find no acceptable fits
at their lower limit (solar composition) despite scanning a larger
range in \amix, but do find solutions for the upper end, ${\rm [Fe/H]}
= +0.10$. These fits may be visualized in the left panel of
Figure~\ref{fig:aycam}, in which each acceptable solution is marked
with a dot and may correspond to a different mixing length parameter
within the limits we explored.  Solutions with $\fov < 0.005$ place
the primary in a rapid stage of evolution, so we have chosen to accept
only values of 0.005 or larger, corresponding to slower phases.  The
full ranges for the primary and secondary are indicated by the
crosshatched region.

\setlength{\tabcolsep}{2pt} 
\setlength{\tabcolsep}{2pt} 

The right panel of Figure~\ref{fig:aycam} displays the allowed
\fov\ ranges when we hold the mixing length parameter fixed at the
value of $\amix_{\sun} = 1.60$ used by \cb\ for both components. This
shrinks the allowed region somewhat, as expected, which now becomes
considerably smaller than reported by \cb, particularly for the
primary. As before, the permitted interval for the secondary is larger
than the primary because it is relatively unevolved.  As done
previously for $\chi^2$~Hya, we considered also solutions with $\amix
= 1.80$, near our value for the Sun. In this case the primary
\fov\ values are confined between 0.010 and 0.020 (squares). The
sections of parameter space allowed by these two choices of \amix\ are
shown by the hatched areas.

\cb\ constrained the \fov\ values for the AY~Cam components to be the
same, although the mass difference is actually non-negligible
\citep[$1.905 \pm 0.040~M_{\sun}$ and $1.709 \pm 0.036~M_{\sun}$, $q =
  0.8972 \pm 0.0032$;][]{Williamon:2004}. If we were to apply the same
constraint and force the two stars to have the same \fov, the ranges
permitted by the observations at a fixed value of \amix\ (as per \cb)
would be even smaller. Thus, under similar conditions as \cb, we find
the degree of overshooting to be considerably better defined for this
system than indicated by those authors.

\subsection{HD 187669}

\begin{figure}
\centering
\begin{tabular}{cc}
\includegraphics[width=4.1cm]{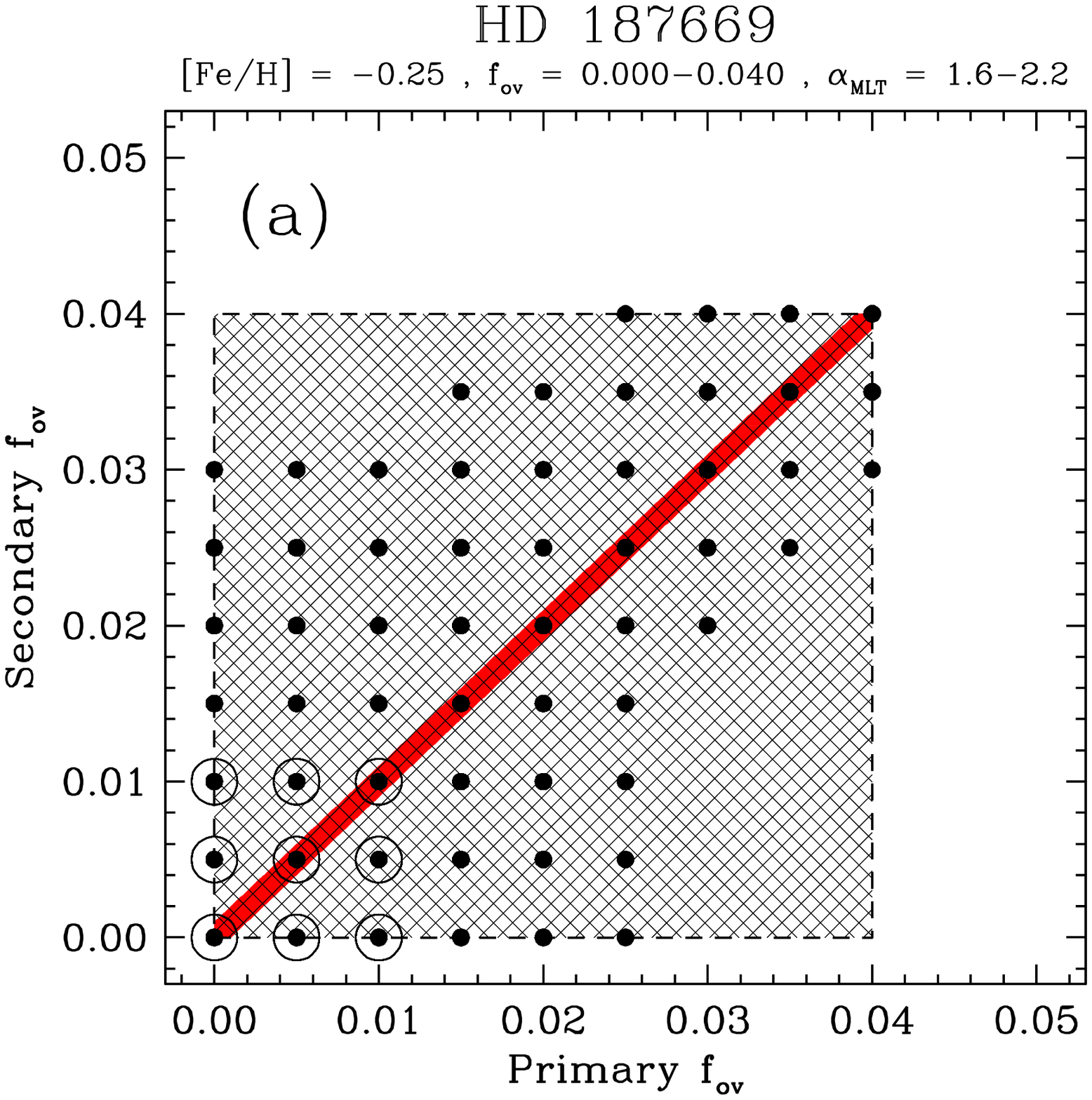} &
\includegraphics[width=4.1cm]{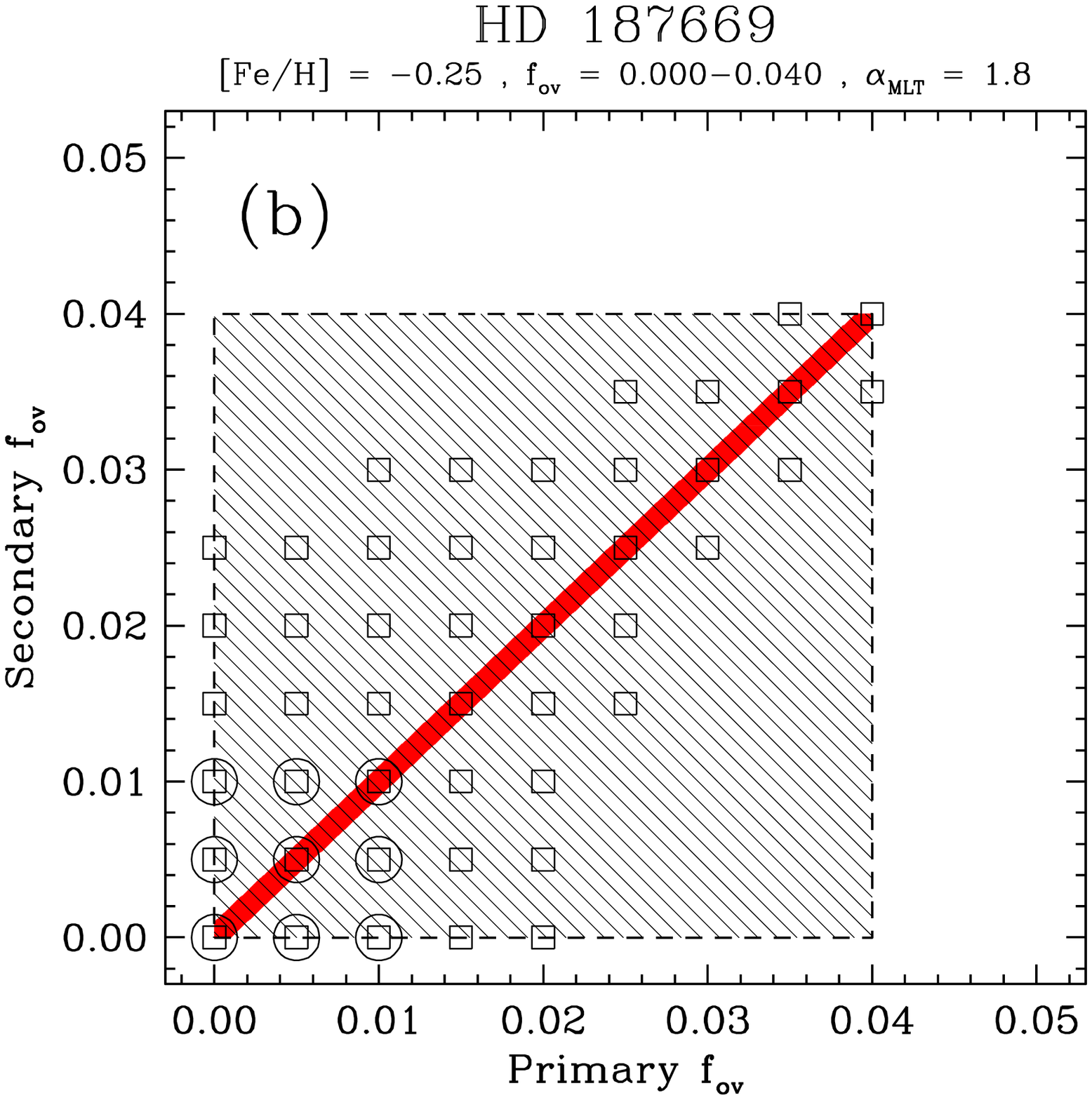}
\end{tabular}
\figcaption{(a) Same as Figure~\ref{fig:ogle227} for HD~187669. (b)
  Acceptable fits for \amix\ set to 1.80 (squares). Open circles in
  both panels mark solutions with \fov\ within the upper limit
  established by \cite{Roxburgh:1992} for stars of this mass (see
  Appendix~\ref{sec:theory}).\label{fig:hd187669}}
\end{figure}

In this case we computed models over the full range in \fov\ specified
by \cb\ (0.000--0.040, with a step of 0.005), and allowed \amix\ to
vary between 1.60 and 2.20 every 0.10. \cb\ held the mixing length
parameter fixed at their solar-calibrated value of $\amix_{\sun} =
1.60$ for both stars. We adopted the same metallicity as \cb, ${\rm
  [Fe/H]} = -0.25$, which is the measured value for this object
\citep{Helminiak:2015}.  The models that match the observations within
errors are marked in Figure~\ref{fig:hd187669}a with circles. We find
no viable solutions for the \amix\ value adopted by \cb, but do find
them for $\amix = 1.80$.  These may be seen in
Figure~\ref{fig:hd187669}b.

\setlength{\tabcolsep}{2pt} 
\setlength{\tabcolsep}{2pt} 

For this binary star the full range of viable \fov\ values (in both
panels of the figure) is as large as that found by \cb\ for both
components, i.e., 0.000--0.040.  Note also that our grids show a
tendency for the good fits to cluster along the diagonal of the area,
consistent with the fact that the two components of HD~187669 have
essentially identical masses \citep[$1.505 \pm 0.004~M_{\sun}$ and
  $1.504 \pm 0.004~M_{\sun}$;][]{Helminiak:2015}.

\vfill
\subsection{BK Peg}

\begin{figure}[!h]
\centering
\begin{tabular}{cc}
\includegraphics[width=4.1cm]{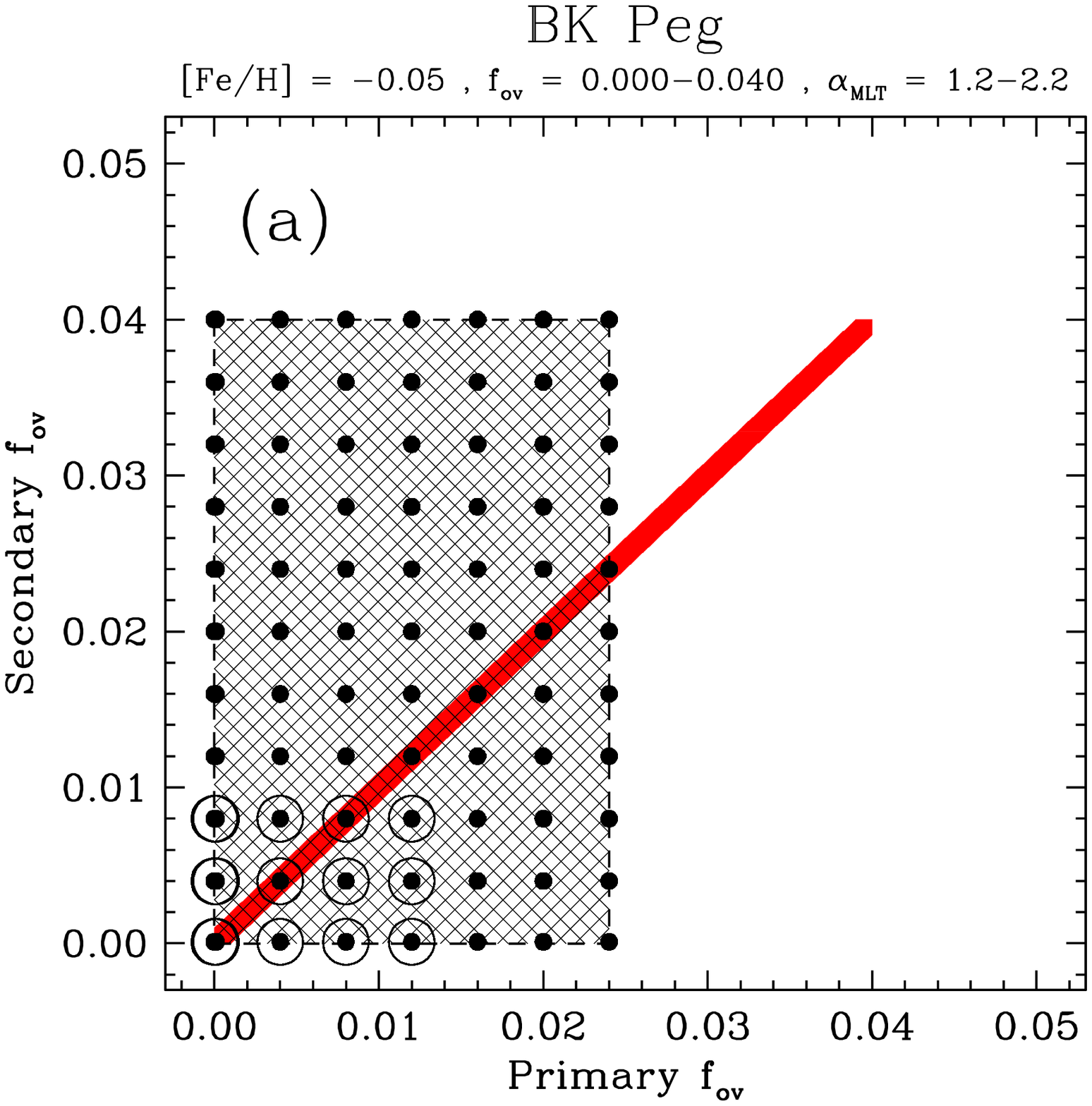} &
\includegraphics[width=4.1cm]{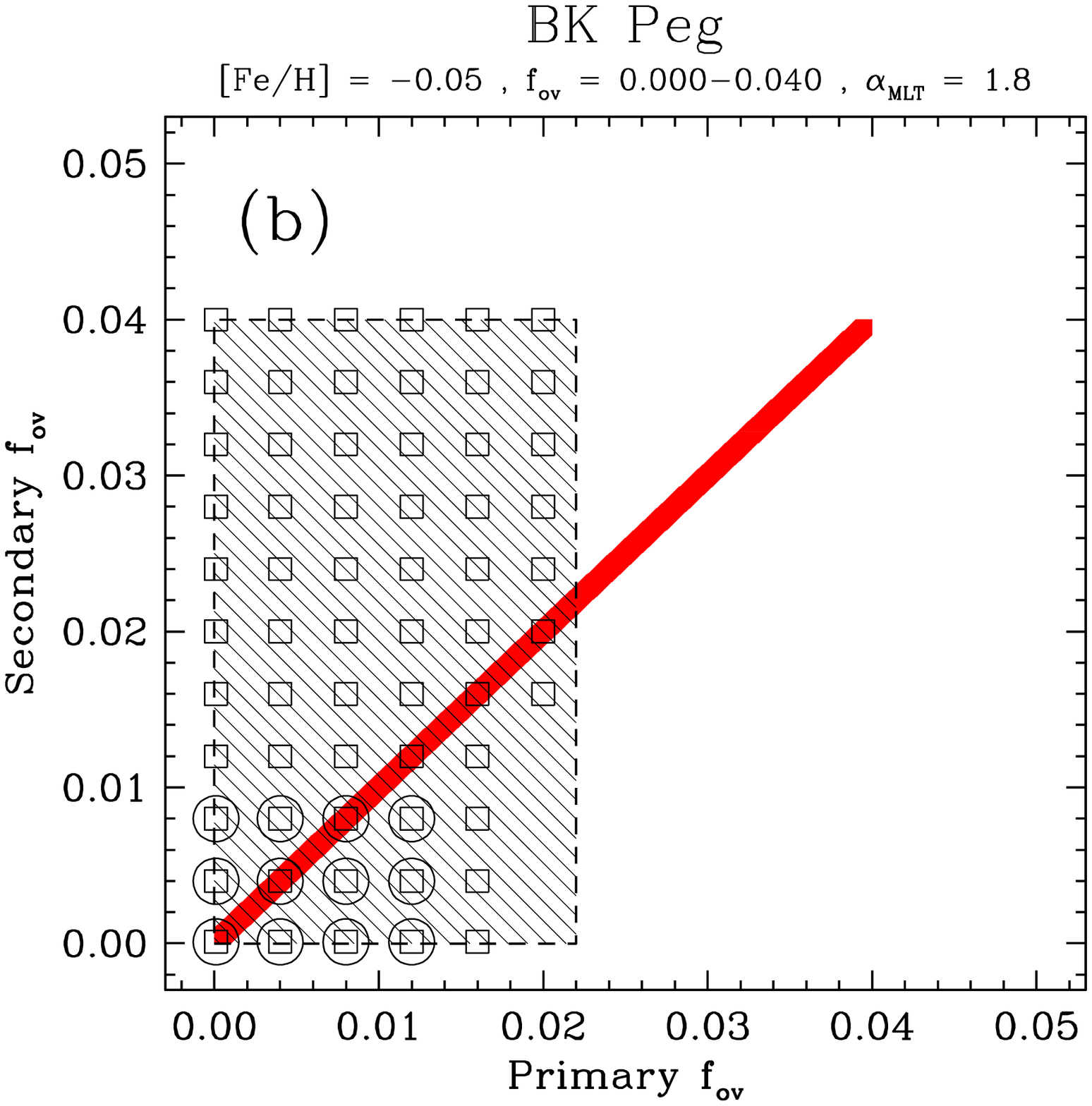}
\end{tabular}
\figcaption{(a) Same as Figure~\ref{fig:ogle227} for BK~Peg.  (b)
  Acceptable fits for \amix\ set to 1.80 (squares). Open circles in
  both panels mark solutions with \fov\ within the upper limit
  established by \cite{Roxburgh:1992} for stars of this mass (see
  Appendix~\ref{sec:theory}). \label{fig:bkpeg}}
\end{figure}

Our grids in \fov, with a fine step size of 0.002, span the same range
as \cb\ (0.000--0.040). For \amix\ we explored values of 1.20--2.20 in
steps of 0.10, which is much wider than the interval of 1.23--1.35
used by \cb. The measured metallicity of BK~Peg is ${\rm [Fe/H]} =
-0.12 \pm 0.07$ \citep{Clausen:2010}. \cb\ explored models in the
[Fe/H] range between $-0.06$ and $+0.05$ that goes beyond the
observational errors. We explored an even wider range between $-0.13$
and $+0.05$, and find acceptable solutions only for the upper limit,
${\rm [Fe/H]} = +0.05$. We display them in Figure~\ref{fig:bkpeg}a,
where it can be seen that the \fov\ values for the primary are better
constrained than reported by \cb. Restricting \amix\ for both stars to
the small interval 1.23--1.35 results in no viable solutions, except
combinations with one star in that range and the other far outside it
($\amix > 1.70$). Mixing length parameters as low as those considered
by \cb\ are rather unusual. In Figure~\ref{fig:bkpeg}b we show the
good fits with $\amix = 1.80$ for both stars. The parameter space for
\fov\ becomes marginally smaller than before in the case of the
primary star.

\setlength{\tabcolsep}{2pt} 
\setlength{\tabcolsep}{2pt} 

We point out, finally, that \cb\ constrained the \fov\ values for the
BK~Peg components to be identical despite noting that the mass
difference is actually non-negligible \citep[$1.414 \pm
  0.007~M_{\sun}$ and $1.257 \pm 0.005~M_{\sun}$, $q = 0.889 \pm
  0.002$;][]{Clausen:2010}, and that this could imply the ``correct''
values of \fov\ should be different if \fov\ depends on mass. Here we
have not made that assumption.

\vfill
\subsection{Summary of results}
\label{sec:summary}

The figures in the preceding subsections (right panels) show that for
most systems we obtain allowable ranges in \fov\ for the primary
and/or secondary that are consistently smaller than reported by \cb,
even though we did not constrain the overshooting to be the same for
the two stars, as they did. Doing so would shrink our permitted
\fov\ ranges even more (see below).  We believe that in some cases
(LMC-562.05-9009, BK~Peg) the \cb\ intervals for \fov\ are larger
partly because they allowed the mixing length parameters to reach
values beyond what is typically considered realistic for similar
stars, whereas we find perfectly acceptable fits without the need for
such extreme values. In the case of OGLE-LMC-ECL-26122 the
metallicities allowed by \cb\ also deviated considerably from the
measured composition, whereas we have always kept the models within
the observational errors in [Fe/H], when available, or explored the
same range as \cb\ otherwise. The systems for which we tend to agree
the most are ones in which \cb\ did not fix the mixing length
parameter to their solar-calibrated value (OGLE-LMC-ECL-CEP-0227,
LMC-562.05-9009, OGLE-LMC-ECL-26122).

While all our calculations have included the effects of microscopic
diffusion, it is not clear whether or not \cb\ did the same in their
work. The influence of diffusion can be very important, particularly
for low and intermediate mass stars. It enhances the heavy element
content of the core, increasing its size as the star evolves. Thus, to
some degree it plays a similar role as overshooting, and can affect
the determination of \fov\ \citep[see, e.g.,][]{Michaud:2004,
  Deheuvels:2016}.  If \cb\ did not include diffusion, then there
would be a tendency for them to allow higher values of \fov\ in their
fits, such as they report.  Additionally, diffusion alters the surface
abundances (i.e., the abundances accessible to observation) to a
degree that depends on metallicity and evolutionary state \citep[see,
  e.g.,][]{Dotter:2017, Deal:2018}. Enforcing the measured (surface)
abundances when fitting models while at the same time ignoring
diffusion can therefore bias the results, including the inferred
\fov\ and \amix\ values.

We emphasize here again that our calculations do not take into account
the influence of stellar rotation, which can also lead to extra mixing
and is therefore degenerate with the effect of overshooting, to some
extent \citep[see, e.g.,][]{Ekstrom:2012}. Including rotation in the
modeling would tend to result in lower values of \fov. However, we do
not expect rotation to change the overall shape of any dependence
there may be between the overshooting parameter and mass.

\section{Assessment of the constraint on \fov\ from double-lined
eclipsing binaries}
\label{sec:fovstrength}

\begin{figure}
\epsscale{1.15}
\plotone{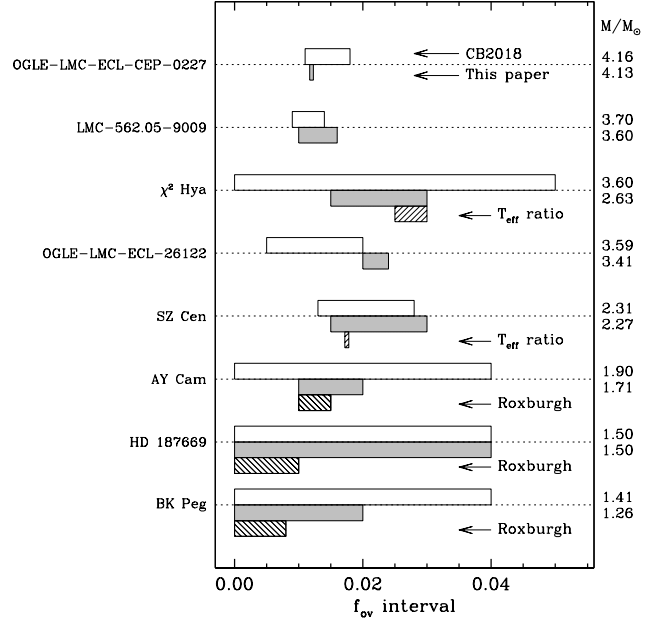}

\figcaption{Comparison between the acceptable ranges for the
  overshooting parameter from \cb\ (white rectangles) and from our own
  grids (gray rectangles) assuming the same values of \fov\ for the
  primary and secondary, and under closely similar conditions
  regarding \amix. Systems are in the same order as discussed
  previously, and primary and secondary masses in solar units are
  indicated on the right.  The hatched rectangles for $\chi^2$~Hya and
  SZ~Cen represent the much reduced ranges we obtain by adding the
  constraint from the temperature ratio (see previous section). For
  the three lower-mass systems AY~Cam, HD~187669, and BK~Peg the
  hatched areas result from considering the maximum plausible values
  of \fov\ from theory \citep{Roxburgh:1992} (see
  Appendix~\ref{sec:theory}). \label{fig:cb2018}}
\end{figure}

If we now restrict \fov\ (along with \amix) to be the same for the two
components of each binary to place our results on the same footing as
those of \cb, the \fov\ intervals become smaller in several cases.
Figure~\ref{fig:cb2018} displays a comparison of our
\fov\ uncertainties (gray shaded rectangles) with theirs (open
rectangles) for all eight DLEBs in their sample. In one case
(OGLE-LMC-ECL-CEP-0227) our grid search returns a single acceptable
value of \fov. On the other hand, for LMC-562.05-9009, SZ~Cen, and
HD~187669 we find good fits over essentially the same ranges as \cb.
There are, however, additional considerations that should not be
ignored and that rule out some of the fits, making the formal
uncertainties in \fov\ smaller.

One of those considerations is the tight constraint on the temperature
ratio that is available for $\chi^2$~Hya and SZ~Cen, which \cb\ did
not use. The impact of this was discussed in the previous section, and
is indicated again in Figure~\ref{fig:cb2018} with an arrow for these
two systems. It shows that for $\chi^2$~Hya the viable range in
\fov\ is reduced to one tenth the size reported by \cb, and for SZ~Cen
it is confined to a single point from our grid.

In five of the eight DLEBs studied by \cb\ the authors claimed the
observations are formally consistent with values of \fov\ exceeding
0.025, and even as high as 0.050 in the case of $\chi^2$~Hya. Aside
from the fact that there is no credible empirical evidence for such a
large degree of overshooting as this for stars in the mass range of
this sample ($\sim$1.2--4~$M_{\sun}$), we believe a more serious issue
is that these extreme \fov\ values run counter to common sense limits
grounded in theory. For stars under about 2.0~$M_{\sun}$ that have
small convective cores it is well known that typical values of the
overshooting parameter coupled with a direct dependence between the
expansion of the core and the pressure scale height would lead to core
sizes that become unrealistically large for such stars because of the
sharp increase in the pressure scale height \citep[see,
  e.g.,][]{Wuchterl:1998, Woo:2001, Pietrinferni:2004,
  Demarque:2004}. A more detailed discussion of this issue and its
consequences may be found in the Appendix.  \cite{Roxburgh:1992}
established that for stars with small cores the extension due to
overshoot cannot exceed about 18\% of the size of the classical core
set by the Schwarzschild criterion \citep[see also][]{Woo:2001}. This
maximum core size effectively limits \fov\ to values much smaller than
those allowed by \cb, and matters the most for the three lowest-mass
stars in the sample, AY~Cam, HD~187669, and BK~Peg. These are
precisely the ones that led \cb\ to claim a lack of evidence for a
dependence of \fov\ on mass. Referring back to
Figures~\ref{fig:aycam}--\ref{fig:bkpeg} for these three systems, we
have highlighted with large open circles the grid points that satisfy
the Roxburgh upper limit. The corresponding ranges allowed for
\fov\ have also been transcribed to Figure~\ref{fig:cb2018}, and are
indicated with an arrow. They are between four and eight times smaller
than the \cb\ uncertainties.

Thus, based on our grid experiments and the discussion above, we feel
that the general conclusion reached by \cb\ regarding the lack of
utility of DLEBs for estimating \fov\ is overly pessimistic, and
perhaps somewhat misleading. From their work it is apparent that their
approach to the issue of estimating error bars for \fov\ was strictly
statistical in nature, with no other consideration. The situation is
more nuanced, as we have shown, and there are other empirical as well
as theoretical constraints that should be taken into account if the
uncertainties are to be physically realistic. We have been mindful of
these constraints in our previous work, even if not stated explicitly,
particularly of the theoretical upper limit on the size of the
convective cores for lower-mass stars that seems fairly obvious to us.

However, it is clear --- and in this we agree with \cb\ --- that the
constraint on \fov\ is much weaker for some of the stars, particularly
the secondaries of $\chi^2$~Hya, AY~Cam, and BK~Peg because they are
less evolved, as well as both components in HD~187669 for the reasons
mentioned by \cb.  For the unevolved stars this is hardly unexpected,
of course, and we have indicated as much in our previous papers. We
concur also that while the strength of the constraint on \fov\ can
vary significantly from system to system, there is value in an
approach such as ours that draws on a large sample of binaries, adding
together whatever information each system is able to contribute for
the purpose of investigating how the overshooting parameter may depend
on mass.

We point out also that in our previous studies the typical error bars
adopted for the \fov\ estimates from the eclipsing binaries were 0.003
for giants and 0.004 for dwarfs, which we still believe to be quite
reasonable when all constraints are taken into account, as described
above.  Nevertheless, based on the much more detailed examination of
the eight systems in the present study as well as additional
experiments with new binaries presented below, we consider it prudent
to adjust our earlier error estimates to 0.004 for giants and 0.006
for unevolved stars so as to be more conservative. We adopt these
uncertainties also in the following.

\section{New \fov\ determinations}
\label{sec:newbinaries}

Since our most recent study the properties of several of the DLEBs
studied in Paper~II have been updated by \cite{Graczyk:2018}, in some
cases significantly. Additionally, we have identified 12 more
well-measured detached binary systems (mass and radius uncertainties
less than $\sim$2\%) in which one or both components are sufficiently
evolved to be useful for estimating \fov. Most have metallicity
estimates. Almost all the new binaries are giants, and in five cases
one or both stars are less massive than 2~$M_{\sun}$, and are
therefore especially valuable for investigating the dependence of
\fov\ on mass. We have also added $\chi^2$~Hya to the list, which we
investigated in detail above as part of the \cb\ sample. This system
had been dropped from our previous studies because we had been unable
to obtain a good fit to the observations at sufficiently similar ages
for the two components using a different set of models (see Paper~I),
but we have now been able to do so with MESA.

The revised physical parameters for the nine DLEBs from
Paper~II\footnote{Note that in revising the properties of
    OGLE-LMC-ECL-26122 \cite{Graczyk:2018} have also changed its
    designation to OGLE-LMC-SC9-230659; for consistency we continue to
    use the old name here.} and for the 13 new binaries
including $\chi^2$~Hya are collected in Table~\ref{tab:combinedtable}.
Each block of the table has the systems sorted in decreasing order of
the primary mass. In two of the new systems, KIC~10031808 and
KIC~9246715 \citep{Helminiak:2019}, the spectroscopic abundance
analysis also yielded a measure of [$\alpha$/Fe], which we have
accounted for below in fitting the evolutionary tracks.

\begin{deluxetable*}{lccccc}
\tabletypesize{\scriptsize}
\tablewidth{0pc}
\tablecaption{Revised parameters and new binaries systems.\label{tab:combinedtable}}
\tablehead{
\colhead{Name} & 
\colhead{Mass ($M_{\sun}$)} & 
\colhead{Radius ($R_{\sun}$)} & 
\colhead{$T_{\rm eff}$ (K)} & 
\colhead{[Fe/H]} &
\colhead{Source}
}
\startdata
\multicolumn{6}{c}{Revised parameters for binaries in our Paper~II sample} \\ [0.5ex]
\noalign{\hrule} \\ [-1.5ex]
OGLE-LMC-ECL-06575  & $4.167 \pm 0.022$ & $43.93 \pm 0.43$\phn & $4920 \pm  80$\phn\phn & $-0.46 \pm 0.10$\phs & 1 \\ 
                    & $3.989 \pm 0.026$ & $46.75 \pm 0.43$\phn & $4645 \pm  60$\phn\phn &                      &   \\ 
OGLE-LMC-ECL-26122\tablenotemark{a}  &  $3.598 \pm 0.038$  &  $32.83 \pm 0.22$\phn   &  $5000 \pm  70$\phn \phn &  $-0.24 \pm 0.11$\phs  & 1 \\
                    &  $3.429 \pm 0.030$  &  $23.40 \pm 0.31$\phn   &  $5030 \pm 100$\phn      &                        &   \\
OGLE-LMC-ECL-01866  & $3.550 \pm 0.031$ & $47.11 \pm 0.50$\phn & $4495 \pm  60$\phn\phn & $-0.49 \pm 0.17$\phs & 1 \\ 
                    & $3.560 \pm 0.020$ & $27.79 \pm 0.52$\phn & $5300 \pm  80$\phn\phn &                      &   \\ 
OGLE-LMC-ECL-10567  & $3.333 \pm 0.029$ & $24.60 \pm 0.29$\phn & $5065 \pm 100$\phn     & $-0.70 \pm 0.10$\phs & 1 \\ 
                    & $3.184 \pm 0.026$ & $36.64 \pm 0.25$\phn & $4715 \pm  75$\phn\phn &                      &   \\ 
OGLE-LMC-ECL-09114  & $3.304 \pm 0.023$ & $26.33 \pm 0.34$\phn & $5230 \pm  60$\phn\phn & $-0.38 \pm 0.12$\phs & 1 \\ 
                    & $3.205 \pm 0.025$ & $18.79 \pm 0.37$\phn & $5425 \pm 110$\phn     &                      &   \\ 
OGLE-LMC-ECL-09660  & $2.997 \pm 0.012$ & $44.40 \pm 0.26$\phn & $4685 \pm  95$\phn\phn & $-0.46 \pm 0.10$\phs & 1 \\ 
                    & $2.981 \pm 0.013$ & $23.66 \pm 0.21$\phn & $5250 \pm  65$\phn\phn &                      &   \\ 
OGLE-LMC-ECL-25658  & $2.231 \pm 0.024$ & $21.40 \pm 0.15$\phn & $4840 \pm  70$\phn\phn & $-0.48 \pm 0.13$\phs & 1 \\ 
                    & $2.230 \pm 0.023$ & $27.61 \pm 0.19$\phn & $4720 \pm  75$\phn\phn &                      &   \\
OGLE-LMC-ECL-03160  & $1.802 \pm 0.018$ & $37.42 \pm 0.24$\phn & $4450 \pm  70$\phn\phn & $-0.68 \pm 0.18$\phs & 1 \\ 
                    & $1.792 \pm 0.016$ & $17.03 \pm 0.28$\phn & $4930 \pm 100$\phn     &                      &   \\ 
OGLE-LMC-ECL-15260  & $1.449 \pm 0.018$ & $23.22 \pm 0.43$\phn & $4810 \pm 130$\phn     & $-0.63 \pm 0.12$\phs & 1 \\ 
                    & $1.422 \pm 0.016$ & $42.20 \pm 0.92$\phn & $4420 \pm  85$\phn\phn &                      &   \\ [0.5ex]
\noalign{\hrule} \\ [-1.5ex]
\multicolumn{6}{c}{New binary systems} \\ [0.5ex]
\noalign{\hrule} \\ [-1.5ex]
OGLE-LMC-ECL-13360  &  $4.060 \pm 0.024$  &  $39.46 \pm 0.35$\phn   &  $5085 \pm  80$\phn\phn  &  $-0.30 \pm 0.10$\phs  & 1 \\
                    &  $3.950 \pm 0.024$  &  $30.46 \pm 0.38$\phn   &  $5495 \pm  90$\phn\phn  &                        &   \\
$\chi^2$ Hya        &  $3.605 \pm 0.078$  &  $4.390 \pm  0.039$     & $11750 \pm 190$\phn\phn  &    \nodata             & 2 \\
                    &  $2.632 \pm 0.049$  &  $2.159 \pm 0.030$      & $11100 \pm 230$\phn\phn  &                        &   \\
OGLE-LMC-ECL-21873  &  $3.093 \pm 0.024$  &  $24.67 \pm 0.24$\phn   &  $5055 \pm  80$\phn\phn  &  $-0.28 \pm 0.12$\phs  & 1 \\
                    &  $2.984 \pm 0.021$  &  $20.26 \pm 0.22$\phn   &  $5265 \pm  75$\phn\phn  &                        &   \\
OGLE-LMC-ECL-24887  &  $2.976 \pm 0.045$  &  $17.83 \pm 0.30$\phn   &  $5070 \pm  80$\phn\phn  &  $-0.22 \pm 0.12$\phs  & 1 \\
                    &  $2.747 \pm 0.047$  &  $16.43 \pm 0.26$\phn   &  $5130 \pm  80$\phn\phn  &                        &   \\
OGLE-LMC-ECL-18836  &  $2.858 \pm 0.031$  &  $15.95 \pm 0.25$\phn   &  $5155 \pm 100$\phn      &  $-0.40 \pm 0.10$\phs  & 1 \\
                    &  $2.784 \pm 0.036$  &  $30.87 \pm 0.33$\phn   &  $4605 \pm  80$\phn\phn  &                        &   \\
OGLE-LMC-ECL-13529  &  $2.857 \pm 0.016$  &  $17.03 \pm 0.21$\phn   &  $5295 \pm  75$\phn\phn  &  $-0.18 \pm 0.14$\phs  & 1 \\
                    &  $2.810 \pm 0.016$  &  $15.98 \pm 0.22$\phn   &  $5260 \pm  90$\phn\phn  &                        &   \\
V4089 Sgr           &  $2.584 \pm 0.008$  &  $3.959 \pm 0.013$      &  $8433 \pm  97$\phn\phn  &     \nodata            & 3 \\
                    &  $1.607 \pm 0.007$  &  $41.605 \pm 0.016$\phn &  $7631 \pm 105$\phn      &                        &   \\
OGLE-LMC-ECL-09678  &  $2.549 \pm 0.031$  &  $30.60 \pm 0.28$\phn   &  $4705 \pm  90$\phn\phn  &  $-0.38 \pm 0.16$\phs  & 1 \\
                    &  $2.400 \pm 0.029$  &  $13.77 \pm 0.16$\phn   &  $5230 \pm  80$\phn\phn  &                        &   \\
KIC 9246715        &  $2.1869 \pm 0.0033$ &  $8.49 \pm 0.12$        &  $4890 \pm 50$\phn\phn   &  \phm{\tablenotemark{a}}$+0.01 \pm 0.03$\tablenotemark{b}\phs  & 4 \\
                   &  $2.1598 \pm 0.0032$ &  $8.20 \pm 0.09$        &  $4905 \pm 60$\phn\phn   &                        &   \\
OGLE-LMC-ECL-12669  &  $1.962 \pm 0.030$  &  $23.36 \pm 0.30$\phn   &  $4715 \pm  95$\phn\phn  &  $-0.30 \pm 0.14$\phs  & 1 \\
                    &  $1.843 \pm 0.029$  &  $24.17 \pm 0.34$\phn   &  $4630 \pm  85$\phn\phn  &                        &   \\
OGLE-LMC-ECL-12875  &  $1.858 \pm 0.023$  &  $40.82 \pm 0.32$\phn   &  $4385 \pm 110$\phn      &  $-0.48 \pm 0.15$\phs  & 1 \\
                    &  $1.831 \pm 0.020$  &  $15.62 \pm 0.13$\phn   &  $4845 \pm 100$\phn      &                        &   \\
KIC 10031808        &  $1.798 \pm 0.013$  &  $3.027 \pm 0.014$      &  $6840 \pm 105$\phn      &  \phm{\tablenotemark{a}}$-0.11 \pm 0.08$\tablenotemark{b}\phs  & 4 \\
                    &  $1.741 \pm 0.009$  &  $2.590 \pm 0.020$      &  $7105 \pm 110$\phn      &                        &   \\
OGLE-LMC-ECL-12933  &  $1.516 \pm 0.019$  &  $17.32 \pm 0.22$\phn   &  $4900 \pm 200$\phn      &  $-0.38 \pm 0.13$\phs  & 1 \\
                    &  $1.514 \pm 0.017$  &  $36.41 \pm 0.31$\phn   &  $4470 \pm 150$\phn      &                        &   
\enddata
\tablenotetext{a}{The new designation introduced by \cite{Graczyk:2018} is OGLE-LMC-SC9-230659; we retain the old name for consistency with our earlier work.}
\tablenotetext{b}{KIC~9246715 has a measured [$\alpha$/Fe] $= -0.01 \pm 0.03$, and KIC~10031808 has [$\alpha$/Fe] $= +0.16 \pm 0.06$.}
\tablecomments{
The first line for each system corresponds to the more evolved star.
Sources are:
(1) \cite{Graczyk:2018}; 
(2) \cite{Torres:2010}.
(3) \cite{Veramendi:2015};
(4) \cite{Helminiak:2019}.
}
\end{deluxetable*}

Our new determinations of \fov\ and \amix\ were carried out with the
same MESA models described in Section~\ref{sec:methodology}, following
the same procedure as in our previous studies. Briefly, we made use of
coarse grids of evolutionary tracks calculated for the measured
component masses and spanning a range of \fov\ and \amix\ values.
These grids served to guide subsequent manual adjustments toward the
final estimates of the overshooting and mixing length parameters for
each star, making use of all observational and theoretical
constraints. The values reported here are those giving the smallest
chi-squared value, with the main observables typically being the
absolute radii and temperatures. The ages for the two components in
each system were allowed to differ by no more than 5\%. Adopted
uncertainties in \fov\ are 0.004 for giants and 0.006 for dwarfs (see
above). The \amix\ values have typical errors of 0.20. For full
details we refer the reader to our earlier studies (e.g., Section~3 of
Paper~II, or Section~3 of Paper~III). The results are presented in
Table~\ref{tab:results}, along with the best-fit heavy element
abundance $Z$ and the mean age of each system.

\setlength{\tabcolsep}{8pt}  %
\begin{deluxetable*}{lccccc@{}c}
\tablecaption{Fitted overshooting and mixing length parameters.\label{tab:results}}
\tablehead{
\colhead{} &
\multicolumn{2}{c}{Primary} &
\multicolumn{2}{c}{Secondary} &&
\\  [-0.5ex]
& \multicolumn{2}{c}{------------------------} &
\multicolumn{2}{c}{------------------------} &&
\colhead{Mean age}
\\
\colhead{Name} & 
\colhead{\fov} & 
\colhead{\amix} & 
\colhead{\fov} & 
\colhead{\amix} &
\colhead{\phm{\tablenotemark{a}}$Z$\tablenotemark{a}} &
\colhead{(Myr)}
}
\startdata
\multicolumn{7}{c}{Revised parameters for binaries in our Paper~II sample} \\ [0.5ex]
\noalign{\hrule} \\ [-1.5ex]
OGLE-LMC-ECL-06575  &   0.0160  &  2.20  &  0.0160  &  2.03  &  0.0048 &   151   \\
OGLE-LMC-ECL-26122  &   0.0160  &  1.80  &  0.0160  &  2.05  &  0.0070 &   233   \\
OGLE-LMC-ECL-01866  &   0.0160  &  1.92  &  0.0150  &  1.97  &  0.0070 &   228   \\
OGLE-LMC-ECL-10567  &   0.0155  &  1.95  &  0.0160  &  2.01  &  0.0050 &   251   \\
OGLE-LMC-ECL-09114  &   0.0160  &  2.00  &  0.0180  &  1.70  &  0.0040 &   253   \\
OGLE-LMC-ECL-09660  &   0.0160  &  2.18  &  0.0170  &  2.05  &  0.0040 &   341   \\
OGLE-LMC-ECL-25658  &   0.0160  &  2.09  &  0.0160  &  2.09  &  0.0044 &   786   \\
OGLE-LMC-ECL-03160  &   0.0070  &  1.87  &  0.0070  &  1.98  &  0.0033 &  1102   \\
OGLE-LMC-ECL-15260  &   0.0040  &  2.20  &  0.0040  &  2.09  &  0.0033 &  2070   \\ [0.5ex]
\noalign{\hrule} \\ [-1.5ex]
\multicolumn{7}{c}{Fitted parameters for the new binary systems in this work} \\ [0.5ex]
\noalign{\hrule} \\ [-1.5ex]
OGLE-LMC-ECL-13360  &   0.0160  &  1.80  &  0.0140  &  1.80  &  0.0080 &   174   \\
$\chi^2$ Hya        &   0.0200  &  1.80  &  0.0150  &  1.80  &  0.0120 &   204   \\
OGLE-LMC-ECL-21873  &   0.0165  &  1.85  &  0.0160  &  1.95  &  0.0053 &   332   \\
OGLE-LMC-ECL-24887  &   0.0160  &  2.20  &  0.0160  &  2.30  &  0.0080 &   397   \\
OGLE-LMC-ECL-18836  &   0.0160  &  2.03  &  0.0160  &  1.95  &  0.0053 &   357   \\
OGLE-LMC-ECL-13529  &   0.0160  &  2.08  &  0.0160  &  2.15  &  0.0053 &   378   \\
V4089 Sgr       &   0.0170  &  2.00  &  0.0060  &  2.10  &  0.0160 &   525   \\
OGLE-LMC-ECL-09678  &   0.0150  &  2.00  &  0.0150  &  1.90  &  0.0040 &   549   \\
KIC 9246715         &   0.0140  &  1.80  &  0.0140  &  1.80  &  0.0130 &   824   \\
OGLE-LMC-ECL-12669  &   0.0160  &  2.10  &  0.0180  &  1.95  &  0.0053 &  1124   \\
OGLE-LMC-ECL-12875  &   0.0140  &  2.00  &  0.0140  &  2.00  &  0.0048 &  1181   \\
KIC 10031808    &   0.0140  &  1.90  &  0.0140  &  1.90  &  0.0125 &  1283   \\
OGLE-LMC-ECL-12933  &   0.0045  &  2.20  &  0.0045  &  2.20  &  0.0048 &  1882   
\enddata
\tablenotetext{a}{Bulk initial composition from our fits.}
\end{deluxetable*}
\setlength{\tabcolsep}{6pt}  

\begin{figure}
\includegraphics[height=6.0cm,trim={0 0 0.4cm 0.02cm},clip]{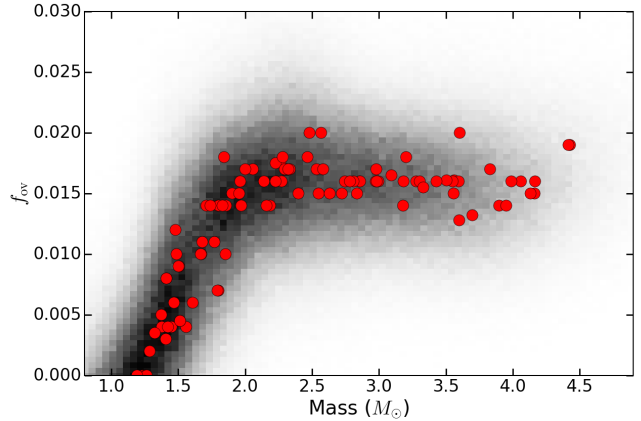}

\figcaption{Overshooting parameter as a function of mass, displayed as
  a heatmap (see text). The 100 semi-empirical measurements of
  \fov\ from this work and our earlier studies are also
  shown. \label{fig:fovmass}}
\end{figure}

With the present determinations along with those reported in Paper~II
(updated here for 9 cases; see Table~\ref{tab:combinedtable})
and Paper~III we now have \fov\ estimates for a total of 50
DLEBs (100 stars), all based on the same MESA models with the
same element mixture \citep{Asplund:2009}. The new and revised
\fov\ values are consistent with the results from our earlier studies,
and support the existence of a dependence with mass. A representation
of this relation that provides a sense of the overall uncertainties is
shown in Figure~\ref{fig:fovmass} in the form of a ``heatmap'', in
which each \fov\ measurement was replaced with a distribution of
10,000 random draws from a Gaussian distribution with $\sigma = 0.004$
or 0.006 depending on whether the star is evolved or not. To achieve
smoothness in the horizontal direction we also perturbed the masses in
the same way with $\sigma = 10$\% of their individual values. Each
pixel in the diagram is colored according to the number of random
draws it contains.  The actual \fov\ measurements are shown as well.

\section{Discussion}
\label{sec:discussion}

The challenging problem of calibrating the convective core
overshooting parameter has been approached in different ways over the
last two or three decades using a variety of empirical data. These
include color-magnitude diagrams of open clusters and associations,
binary stars, rotational velocities of massive stars, and more
recently also asteroseismic constraints.  Due in large part to
observational limitations (measurement errors, sample size, etc.),
many of these determinations have reported estimates of a mean or
typical value of \aov\ or \fov\ for a given population of stars, but
have generally lacked the sensitivity to explore changes as a function
of mass or other stellar properties. As a result, there is conflicting
evidence in the literature: some studies support a variation of the
overshooting parameter with mass, while others find that a constant
value is perfectly adequate.

For example, \cite{Schaller:1992} adopted a fixed overshooting
parameter ($\aov = 0.2$ in the step-function approximation) for stars
of all masses, on the basis of a comparison of their models and
earlier ones with the morphology of the turnoff of several dozen open
clusters \citep[see also][]{Maeder:1981, Mermilliod:1986}.
\cite{Demarque:2004} also used clusters, but favored a gradual
increase in overshooting for masses above $\sim$1.2~$M_{\sun}$
somewhat analogous to what we have found here, with an additional
dependence on metallicity. An independent study by
\cite{Vandenberg:2006} comparing their own models with color-magnitude
diagrams of open clusters reached a qualitatively similar conclusion,
proposing that overshooting increases between about 1.15~$M_{\sun}$
and 1.7~$M_{\sun}$. Further evidence also in the direction of a change
in overshooting with mass was found by \cite{Ekstrom:2012} and
\cite{Georgy:2013} from an investigation of the width of the
terminal-age main-sequence for stars in the range 1.35--9~$M_{\sun}$.
On the other hand, in their recent paper presenting a large grid of
MESA models \cite{Choi:2016} took a different view and adopted a fixed
value of $\fov = 0.016$ for stars of all masses, determined from
isochrone fits to the color-magnitude diagram of M67. On the
asteroseismic side \cite{Aerts:2015} reported seeing no significant
trend of the strength of overshooting with mass for a set of mostly OB
stars, whereas the study of \cite{Deheuvels:2016} did find a hint of a
possible increase with mass over the range $\sim$1.1--1.5~$M_{\sun}$,
although their sample was small.

While it seems undisputed that overshooting is required in most (if
not all) stars above about 2~$M_{\sun}$ \citep[][and
  others]{Moravveji:2015, Moravveji:2016, Valle:2017}, the evidence is
just as compelling that other less massive stars such as the
components of the eclipsing binary AI~Phe ($M = 1.23$ and
1.19~$M_{\sun}$) can only be fit properly \emph{without} overshooting
\citep{Higl:2017}. This imples some sort of transition in the values
of \fov\ or \aov\ between high- and low-mass stars. In fact, in a sign
of a general tendency we see in the literature toward acceptance of
such a transition, many current series of stellar evolution models
(but not all) already incorporate a dependence of overshooting on mass
\citep[e.g.,][]{Demarque:2004, Pietrinferni:2004, Vandenberg:2006,
  Mowlavi:2012, Bressan:2012, Georgy:2013, Spada:2017, Hidalgo:2018},
although the adopted functional form of the trend has so far been
largely arbitrary.

Early studies based on binaries \citep{Schroder:1997, Ribas:2000,
  Claret:2007} also suggested a variation of overshooting with mass,
but were generally afflicted by small samples and/or large
uncertainties. Other more recent binary studies such as those of
\cite{Meng:2014} and \cite{Stancliffe:2015} have not found clear signs
of a change, but also dealt with few systems.  The present
investigation, which continues and expands our previous work in
Paper~II and Paper~III, uses by far the largest sample of binaries
(50) with accurately measured parameters and presents strong evidence
(Figure~\ref{fig:fovmass}) for a gradual change in \fov\ as a function
of mass, with a break point near 2~$M_{\sun}$.

Given the history of claims and counterclaims regarding this issue,
the study of \cb\ comes as no surprise and is in fact welcome, as it
has motivated us to reexamine our own work more critically. They
contend, based on their model fits for a subset of 8 representative
DLEBs they selected from our previous samples, that the formal
uncertainties in the fitted \fov\ values are so large as to make the
more unevolved binaries essentially useless for calibrating
overshooting. Our replication of those experiments with the MESA
models under the same conditions as theirs, to the extent possible,
finds considerably smaller \fov\ uncertainties in many cases,
especially when taking into account the constraint from the
temperature ratio available for some systems, which
\cb\ ignored. Moreover, the implausibly large \fov\ values allowed by
them in several cases are the result of not taking into account
physical arguments that limit the maximum overshooting distance in
stars with small convective cores, specifically, the Roxburgh
criterion. This is particularly relevant below $\sim$2~$M_{\sun}$,
which is precisely the regime in which we (and others) have found that
\fov\ varies with mass (see also Figure~\ref{fig:roxburgh}). The
purely statistical approach of \cb\ for estimating \fov\ is of course
expected to lead to larger formal uncertainties, but is too simplistic
if the goal is to extract overshooting parameters and uncertainties
that make physical sense.  Their assumption of a single \fov\ value in
binaries with components of appreciably different mass ($\chi^2$~Hya,
AY~Cam, BK~Peg) also appears questionable to us, even more so if the
goal is to detect a trend of \fov\ with mass.

Beyond the notion that there is a maximum permissible extension of the
convective core due to overshooting, our current understanding of the
phenomenon does not yet allow us to predict the value of the
overshooting parameter from first principles with any certainty.
Nevertheless, we note that recent theoretical work by
\cite{Jermyn:2018} has led them to postulate that the parameter
depends on mass in much the same way as indicated by the
semi-empirical evidence we have presented here. Their Figure~8 for
\aov\ versus mass bears a striking resemblance to our
Figure~\ref{fig:fovmass} for \fov\ \citep[once the scale factor
  $\aov/\fov \approx 11.4$ is accounted for;][]{Claret:2017}, except
that their peak value of \aov\ seems about 25\% higher than indicated
by the observations. Through similar theoretical arguments they find
an analogous mass dependence for the diffusive parameter \fov\ (see
their Figure~9), although in this case their scale is a factor of
$\sim$5.5 too large.

We mention in closing that a similar trend of overshooting with mass
can be inferred theoretically \citep[and independently of the work
  by][]{Jermyn:2018} from the homology arguments presented by
\cite{Claret:2017}. This follows from Equation~(7) of their Appendix,
which gives an expression for the fractional increase in the mass of
the convective core. Details of this derivation will be published
elsewhere.

\acknowledgments

We thank the anonymous referee for helpful comments that have improved
the manuscript. Dariusz Graczyk is also thanked for alerting us
  to the name change for OGLE-LMC-ECL-26122. The Spanish MEC
(AYA2015-71718-R and ESP2017-87676-C5-2-R) is gratefully acknowledged
for its support during the development of this work.  AC also
acknowledges financial support from the State Agency for Research of
the Spanish MCIU through the ``Center of Excellence Severo Ochoa''
award for the Instituto de Astrof\'isica de Andaluc\'ia
(SEV-2017-0709). GT acknowledges partial support from the NSF through
grant AST-1509375.  This research has made use of the SIMBAD database,
operated at the CDS, Strasbourg, France, and of NASA's Astrophysics
Data System Abstract Service.

\begin{appendix}
\section{Theoretical considerations: upper limits on overshooting}
\label{sec:theory}

\begin{figure}[b!]
\epsscale{0.62}
\plotone{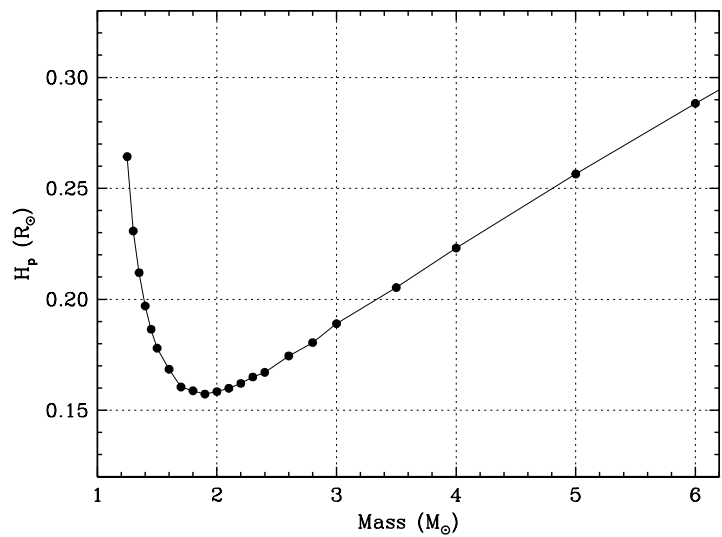}
\figcaption{Pressure scale height at the edge of the standard
  convective core as a function of stellar mass. The relation is based
  on ZAMS models for solar metallicity and $\amix =
  1.80$.\label{fig:Hp}}
\end{figure}

It is helpful to begin our discussion of this issue by considering the
behavior of the key quantity for calculating the overshooting distance
in both the classical step-function approximation (with free parameter
\aov) and the diffusive approximation (\fov), which is the pressure
scale height $H_p$. Using the same MESA code as before, we computed
models over a range of masses and extracted the values of $H_p$ at the
edge of the standard convective core set by the Schwarzschild
criterion. These $H_p$ values are plotted in Figure~\ref{fig:Hp}. For
this illustration we used solar-metallicity zero-age main-sequence
(ZAMS) models\footnote{We define the ZAMS in our models as the stage
  at which the central hydrogen content drops to 99.4\% of its initial
  value.} with $\amix = 1.80$ and no rotation, overshooting, or
microscopic diffusion, and we adopted again the \cite{Asplund:2009}
element mixture.

Interestingly, the $H_p$ trend shows a pronounced minimum at about
1.8~$M_{\sun}$, resulting in values for low-mass stars that are large
and comparable to those of much higher mass stars. This behavior
follows from the expression that defines the pressure scale height at
the radial distance $r$ as:
\begin{equation}
H_p(r) = \frac{P(r)}{\rho(r) g(r)}~,
\end{equation}
where $P(r)$ is the pressure, $g(r)$ the local gravity, and $\rho(r)$
the density. As the convective core becomes very small and $r$
decreases, the result is $H_p \rightarrow \infty$.  Thus, as
Figure~\ref{fig:Hp} shows, a star of 1.25~$M_{\sun}$ has a value of
$H_p$ approximately the same as one of 5~$M_{\sun}$. Therefore,
adopting the same values of \aov\ or \fov\ for these two very
different stars would result in the cores being increased in size by
the same amount.

\begin{figure}
\epsscale{0.6}
\plotone{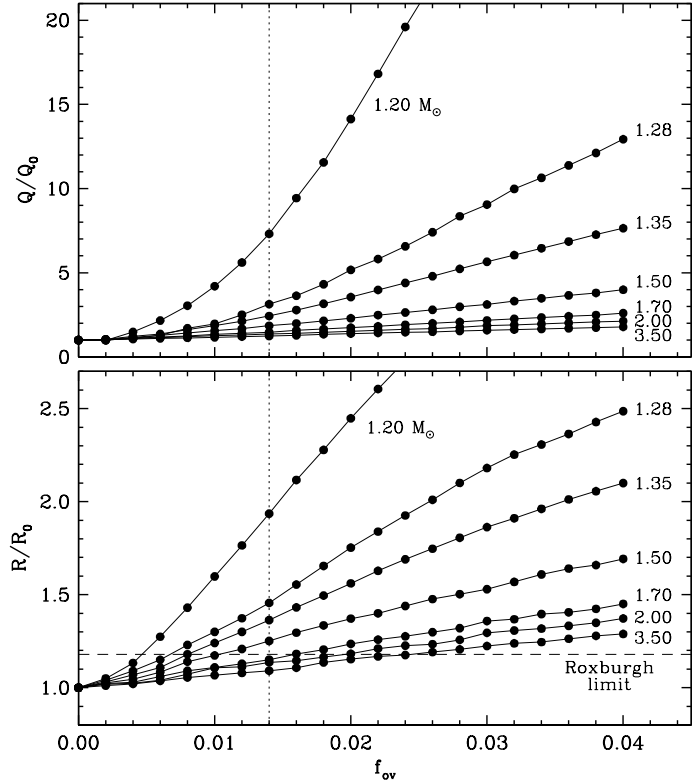}
\figcaption{\emph{Top:} Mass $Q$ of the convective core as a
  function of the overshooting parameter \fov, normalized to the mass
  $Q_0$ of the core without overshooting. Curves are labeled with the
  mass in solar units, and the vertical dotted line marks the
  representative value of $\fov = 0.014$ suggested by \cb.
  Calculations are based on ZAMS models from MESA with $Z = 0.0189$
  and $\amix = 1.80$. \emph{Bottom:} Same as the top panel, for the
  radius $R$ of the convective core normalized to the size $R_0$ of
  the core with no overshooting. The horizontal dashed line follows
  from the Roxburgh criterion for small cores.\label{fig:QR}}
\end{figure}

The top panel of Figure~\ref{fig:QR}, generated with the same models
as above and $Z = 0.0189$, illustrates the change in the mass of the
convective core normalized to its value in the absence of
overshooting, as a function of the overshooting parameter \fov.  For a
reference value of \fov\ such as that indicated by \cb\ ($\fov \approx
0.014$, dotted line), a star of 1.20~$M_{\sun}$ (roughly the lowest
mass at which convective cores appear) would develop a core that is a
full seven times more massive than the classical core, which seems
physically implausible. Even more extreme cores would result for
\fov\ as large as some of the values considered by \cb\ (e.g., 0.040).
For higher mass stars the core enlargements are more modest, and take
on an asymptotic behavior beyond about 1.8~$M_{\sun}$. An equivalent
representation is seen in the bottom panel of Figure~\ref{fig:QR}, now
for the change in the physical size (radius) of the core relative to
that of the classical core with no overshooting.  At the \cb\ value of
$\fov = 0.014$ our 1.20~$M_{\sun}$ star would see its core increased
in size by nearly a factor of two.

This basic physical argument makes it immediately clear that the
formal error bars for \fov\ reported by \cb\ should be reduced
significantly at the upper end, and can hardly reach values as large
as 0.040 for relatively low-mass stars with small cores. This
conclusion is quite aside from the fact that there is no empirical
evidence supporting such large values of \fov\ for stars under about
2~$M_{\sun}$, as pointed out earlier.

Both the step-function and the diffusive formulations of overshooting
are largely ad hoc in nature and lack any predictive power to set
\aov\ or \fov\ from first principles.  A valuable attempt in this
direction was carried out more than 40 years ago by
\cite{Roxburgh:1978}, who established a criterion to estimate the
overshooting distance based on an integral involving the difference
between the luminosities due to nuclear and radiative processes. The
integrand evaluated at the lower and upper limits of the convective
plus overshoot region ($r_1$ and $r_2$, respectively) changes sign at
some intermediate radius $r_{\rm ov}$, permitting a definition of the
overshooting distance as $r_2 - r_{\rm ov}$.

As a device for estimating the overshooting distance itself the
Roxburgh criterion was initially called into question by
\cite{Baker:1987}, but has since been shown to be valid as an
\emph{upper limit} to that distance \citep[see, e.g.,][]{Kuhfuss:1987,
  Zahn:1991, Roxburgh:1992, Canuto:1997}. In particular,
\cite{Roxburgh:1992} estimated that for stars with small cores the
maximum overshooting distance is approximately 0.18 times the size of
the classical core, independently of the details of energy generation
and opacity.  \cite{Woo:2001} arrived at a similar estimate. For
useful discussions about the validity of the Roxburgh criterion, and
more generally about the modeling of convection, we refer the reader
to the reviews by \cite{Maeder:2009} and \cite{Kupka:2017}.

\begin{figure}
\epsscale{0.6}
\plotone{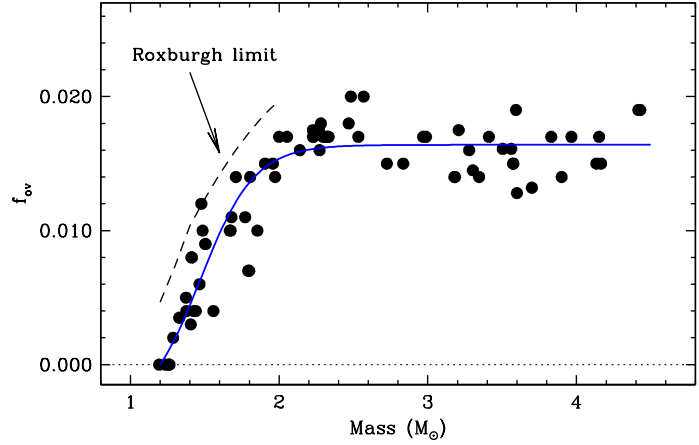}

\figcaption{Previously published estimates of \fov\ from Paper~II and
  Paper~III (37 binary systems) as a function of stellar mass, along
  with the upper limit (dashed line) expected from the criterion of
  \cite{Roxburgh:1978} for stars with small convective cores (see also
  the bottom panel of Figure~\ref{fig:QR}).  The solid line, taken
  from Paper~III, is an approximate representation of the observed
  trend drawn to guide the eye. \label{fig:roxburgh}}

\end{figure}

The \cite{Roxburgh:1992} upper limit to the extension of the core is
shown in bottom panel of Figure~\ref{fig:QR} as a dashed line. This
naturally sets a maximum value of \fov\ for stars with small
convective cores.  We illustrate this more directly in
Figure~\ref{fig:roxburgh}, which reproduces our previous estimates of
\fov\ from \cite{Claret:2017, Claret:2018}. The Roxburgh limit is seen
to be consistent with the semi-empirical determinations of \fov.

This upper limit from theory is most relevant for the three binaries
in the \cb\ sample with the lowest mass: AY~Cam, HD~187699, and
BK~Peg.  For these three systems we have marked in
Figures~\ref{fig:aycam}--\ref{fig:bkpeg} with an open circle the
solutions with \fov\ values for the primary and secondary that satisfy
the Roxburgh criterion (or slightly exceed it, to be conservative).
This constraint reduces the range of allowable values quite
significantly, and should not be ignored.

\end{appendix}

\end{document}